\documentclass[times]{weauth}
\usepackage{moreverb}
\usepackage{subfigure}
\usepackage{graphicx}
\usepackage{dcolumn}
\usepackage{amssymb,amsmath}
\usepackage{epstopdf}
\usepackage{placeins}
\usepackage{color}
\usepackage{epstopdf}
\usepackage{citesort}

\def\volumeyear{2016}

\begin{document}

\abovedisplayskip=5pt
\abovedisplayshortskip=5pt 
\belowdisplayskip=5pt
\belowdisplayshortskip=5pt

\runningheads{Stevens, Gayme, Meneveau}{CWBL model: applications and comparisons with field and LES data}
\articletype{\noindent "This is the peer reviewed version of the following article: \textbf{Stevens, R. J. A. M., Gayme, D. F., and Meneveau, C. (2016) Generalized coupled wake boundary layer model: applications and comparisons with field and LES data for two wind farms. Wind Energ., 19: 2023-2040}, which has been published in final form at \textbf{http://dx.doi.org/10.1002/we.1966}. This article may be used for non-commercial purposes in accordance with Wiley Terms and Conditions for Self-Archiving." \nobreak}
\title{Generalized coupled wake boundary layer model: applications and comparisons with field and LES data for two wind-farms}
\author{Richard J.A.M. Stevens$^{1,2}$, Dennice F. Gayme$^{1}$, and Charles Meneveau$^{1}$}
\address{$^1$ Department of Mechanical Engineering, Johns Hopkins University, Baltimore, Maryland 21218, USA\\
$^2$ Department of Physics, Mesa+ Institute, and J.\ M.\ Burgers Centre for Fluid Dynamics, University of Twente, 7500 AE Enschede, The Netherlands}
\corraddr{r.j.a.m.stevens@utwente.nl}

\begin{abstract}
We describe a generalization of the Coupled Wake Boundary Layer (CWBL) model for wind-farms that can be used to evaluate the performance of wind-farms under arbitrary wind inflow directions whereas the original CWBL model (Stevens {\it et al.}, J.\ Renewable and Sustainable Energy {\bf 7}, 023115 (2015)) focused on aligned or staggered wind-farms. The generalized CWBL approach combines an analytical Jensen wake model with a ``top-down" boundary layer model coupled through an iterative determination of the wake expansion coefficient and an effective wake coverage area for which the velocity at hub-height obtained using both models converges in the ``deep-array" portion (fully developed region) of the wind-farm. The approach accounts for the effect of the wind direction by enforcing the coupling for each wind direction. Here we present detailed comparisons of model predictions with LES results and field measurements for the Horns Rev and Nysted wind-farms operating over a wide range of wind inflow directions. Our results demonstrate that two-way coupling between the Jensen wake model and a ``top-down" model enables the generalized CWBL model to predict the ``deep-array" performance of a wind-farm better than the Jensen wake model alone. The results also show that the new generalization allows us to study a much larger class of wind-farms than the original CWBL model, which increases the utility of the approach for wind-farm designers.
\end{abstract}


\maketitle

\section{Introduction} \label{Section_Introduction}
The geometry and relative turbine positioning in a wind-farm greatly affect the overall power output. This fact has led to a great deal of research into tools for analyzing and improving wind-farm designs. One approach that has been successful in generating accurate wind-farm performance predictions is the use of large eddy simulations (LES) \cite{meh14}. LES have been used to study the performance of existing wind-farms such as Horns Rev \cite{por13} and Lillgrund \cite{chu12,chu12b,lee12,arc13b}. In addition, LES have been used to study the effects of the turbine positioning \cite{ste14b,wu13}, spacing \cite{ste14f,yan12}, and hub-height \cite{arc13b} on the average power output of wind-farms.

While LES is very useful for model development and validation, it is not a practical tool for the design and optimization of individual wind-farms, due to its high computational cost. Instead industry uses wind-farm design tools that employ less computationally expensive methods to evaluate the performance of a specific wind-farm based on a number of different design choices and operating conditions. Reynolds Averaged Navier Stokes (RANS) type models are sometimes used to predict the performance of wind-farm designs \cite{pij06,eec10,sch12,bro12,has09,sch09,bea12,xue00,san11}. While less computationally intensive than LES, RANS is still relatively expensive since it requires solving partial differential equations.  A class of significantly less expensive, but more approximate, methodologies are so-called analytical models that do not require solution of differential equations. Two analytical approaches are commonly used for wind-farm evaluations. The first approach employs wake (``bottom-up'') models \cite{lis79,jen83,kat86,ain88,lar88,cho13,pen14b,pen13b,bas14,nyg14} to estimate the power output of wind-farms. These types of analytical wake models, of which the Jensen model is the most well-known example, have been used in several studies examining layout optimization of wind-farms \cite{mar08,ema10,kus10,saa11}, and are built into a number of commercial packages that are used to predict wind-farm performance. These models describe the power output in the entrance region of a wind-farm well. However, such models do not explicitly include coupling with the vertical structure of the atmospheric boundary layer (ABL), which becomes relevant for very large wind-farms, the effects of this coupling are sometimes referred to as the ``deep-array effect''. In the fully developed wind-farm region, where the turbine power production as function of the downstream direction becomes constant, additional complexities arise due to the vertical structure of the ABL and the associated wake-atmosphere coupling that is not typically captured by wake models. We remark that Nygaard \cite{nyg14} shows that a simple Jensen model, which is similar to that implemented in WAsP (Wind Atlas Analysis and Application Program) but without ``image wakes", can predict the power degradation data from the London Array, Walney and Anholt wind-farms well. That work also states that he did not find evidence of a ``deep-array effects". Stevens {\it et al.} \cite{ste14g} have also shown that in some conditions the Jensen model yields good predictions for particular wind-farm configurations. However, as is shown in figure 1, the Jensen model predictions do not compare well with LES results for very long staggered wind-farms. Moreover some cases shown by Nygaard \cite{nyg14} did show marked differences between data and the Jensen wake model. Wu and Port\'e-Agel \cite{wu15}, Barthelmie {\it et al.} \cite{bar09b}, and Son {\it et al.} \cite{son14} also state that wake models can have difficulty predicting ``deep-array effects".
 
The coupling with the vertical ABL structure can be captured by so called single column or ``top-down'' models \cite{fra92,fra06,cal10,ste14c}, in which the turbines are seen as roughness elements. In this framework, the average velocity profile at hub-height can be obtained based on the assumption of the existence of two logarithmic regions, one above the turbine hub-height and one below. However, ``top-down'' models do not capture the effect of the relative turbine positioning, which is an important design parameter even in the fully developed region of the wind-farm \cite{ste14f}. 

\begin{figure}
\centering
\subfigure{\includegraphics[width=0.99\textwidth]{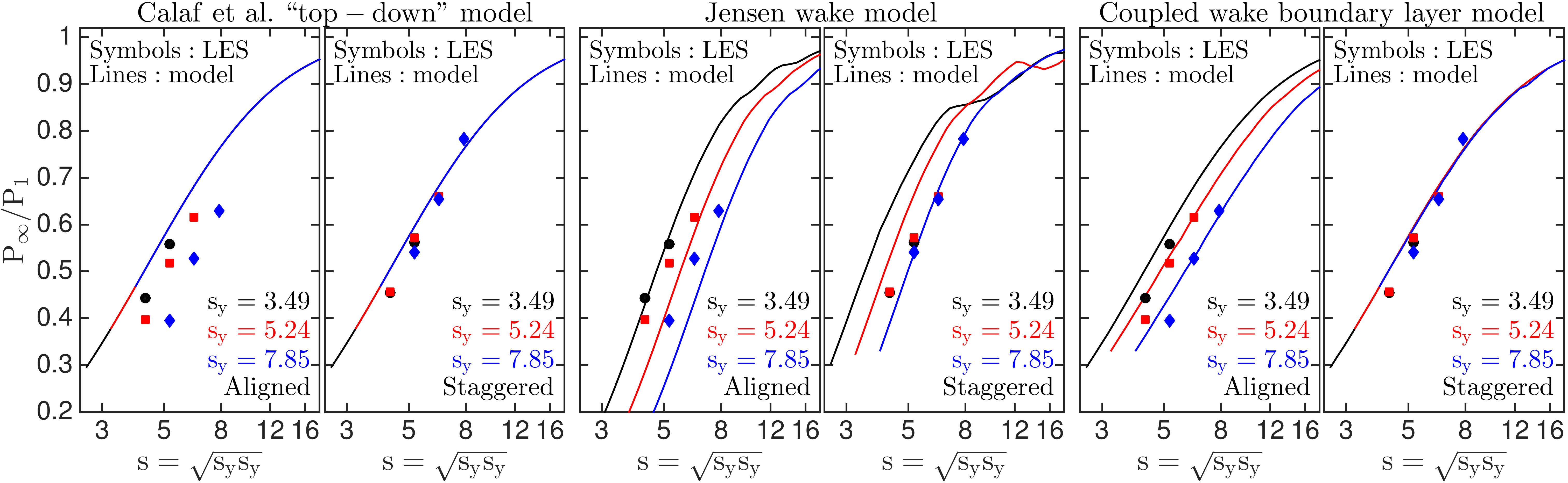}}
\caption{Predictions of the power output ratio $\mathrm{P_\infty}/\mathrm{P_1}$ in the fully developed region for aligned and staggered wind-farms obtained from the Calaf {\it et al.} ``top-down" \cite{cal10,men12,ste14g} model, the Jensen wake model \cite{jen83,kat86}, and the CWBL model compared to LES results. Here $\mathrm{P_1}$ is taken to be the power output of turbines in the first row and $\mathrm{P_\infty}$ the power output in the fully developed region of the wind-farm, which is the region where the power output of the turbines becomes constant as function of the streamwise position. In practice, for these LES results, this is taken to be the average power output of the $8^\mathrm{th}$ or further downstream turbine rows, see details in Refs.\ \cite{ste13,ste14b,ste14f}. These results are plotted as a function of the geometric mean turbine spacing $s=\sqrt{\mathrm{s_x}\mathrm{s_y}}$, where $\mathrm{s_x}$ and $\mathrm{s_y}$ indicate the non-dimensional (in terms of the turbine diameter $D$) streamwise and spanwise distance between the turbines. Note that only the CWBL model captures the trend observed in LES data for both aligned and staggered wind-farms. This figure summarizes the results of Ref.\ \cite{ste14g}.}
\label{figure1}
\end{figure}

The Coupled Wake Boundary Layer (CWBL) model \cite{ste14g} combines the classical Jensen wake model \cite{jen83,kat86,pen14b} and the Calaf {\it et al.} \cite{cal10,men12} ``top-down'' model through a novel two-way coupling. Initial comparisons with data \cite{ste14g} showed that the resulting model provides improved predictions over either of its constitutive analytical models for aligned and staggered wind-farms. The wake model part of the CWBL model ensures that the effect of the relative positioning of the turbines is represented, while the interaction with the ABL in the fully developed region of the wind-farm is captured by the ``top-down'' portion of the model. Figure \ref{figure1} compares the wind-farm performance in the fully developed region of aligned and staggered wind-farms predicted by the Jensen, ``top-down'', and CWBL models with corresponding LES results \cite{ste14g}. This figure clearly shows that the CWBL model gives improved predictions over the wake (Jensen) and ``top-down'' (Calaf {\it et al.}) modeling approaches as only the CWBL model captures the main trends as function of the geometric mean turbine spacing $s=\sqrt{\mathrm{s_x}\mathrm{s_y}}$, where $\mathrm{s_x}$ and $\mathrm{s_y}$ indicate the non-dimensional (in terms of the turbine diameter $D$) streamwise and spanwise distance between the turbines. Here we use the geometrical mean turbine density $s$ in order to provide results for different spanwise $\mathrm{s_y}$ and streamwise $\mathrm{s_x}$ spacings in one plot. As shown by Stevens {\it et al.} \cite{ste14g}, the improved predictions are a result of the two-way coupling between the wake and ``top-down" models and therefore similar performance improvements appear difficult to realize with models that use only one-way coupling between the wake and ``top-down'' modeling approaches, see for example Refs.\ \cite{fra06,pen14b,pen13b,yan15b}.

The CWBL model framework is modular with the two-way coupling between the ``bottom-up" (wake model) and ``top-down" modeling approaches being the crucial component. In this work we selected the ``simplest possible" implementation, which uses a Jensen wake model with the superposition interactions accounted for through the addition of the squared velocity deficits, see details in section \ref{section_jensen}, coupled with the Calaf {\it et al.} \cite{cal10} ``top-down" model, which captures the wind-farm and ABL interaction. Due to the modular approach, the CWBL model can be adapted to use more detailed wake or ``top-down" models. For example the Ainslie \cite{ain88}, Larsen \cite{lar88}, or EPFL \cite{bas14} wake models can be substituted in order to model the wind turbine wakes in greater detail than the Jensen model. Similarly, one can change the model for the wake-wake superposition to, for example, one of the alternatives discussed by e.g.\ Larsen {\it et al.} \cite{lar13} in their validation of the dynamic wake meandering model. In addition, one could also adjust the wake expansion rate that is used in the far wake, see for example the work by Frandsen {\it et al.} \cite{fra06}. We consider these aspects as important topics for future research that can benefit from ongoing developments of improved wake models. Here we continue with the analysis of the CWBL approach in its basic form.

\begin{figure}
\centering
\subfigure{\includegraphics[width=0.80\textwidth]{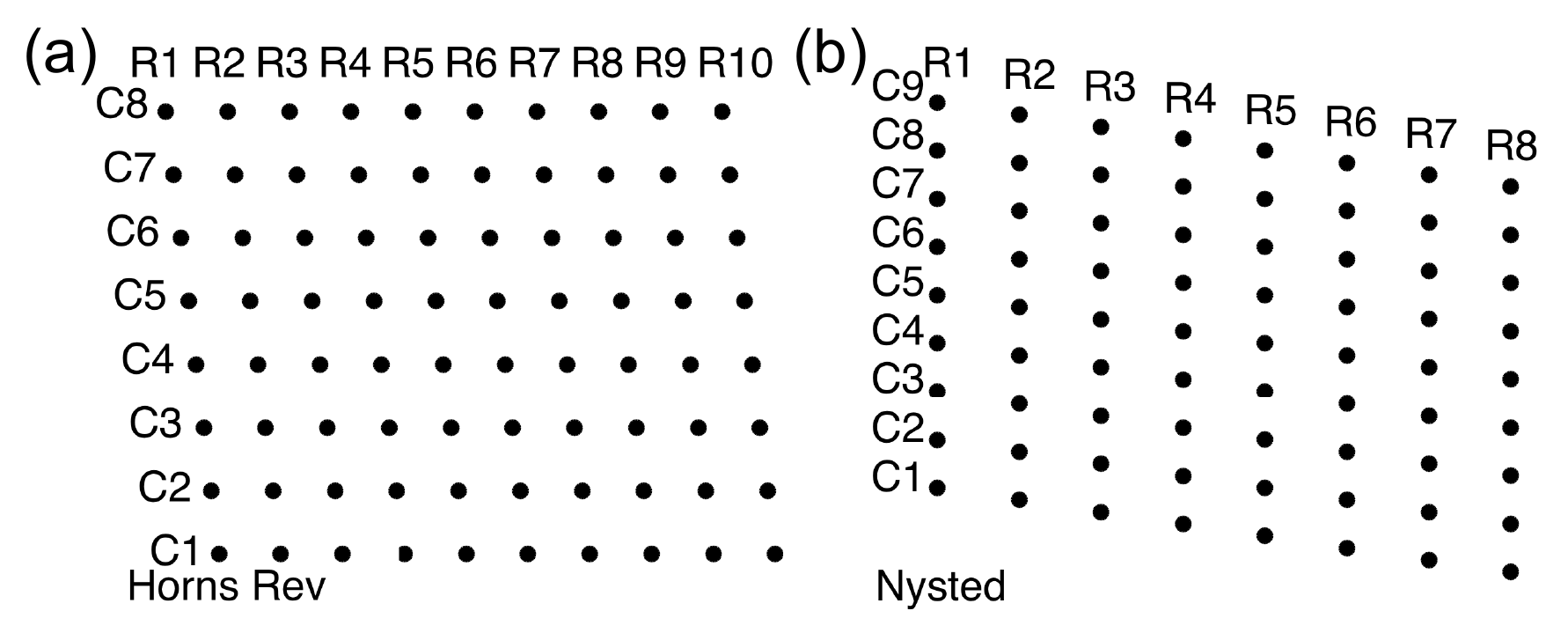}}
\caption{Turbine arrangement of the wind-farms at (a) Horns Rev and (b) Nysted showing the numbering of the rows and columns.}
\label{figure2}
\end{figure}

In this paper we extend the CWBL analytical modeling approach, to more general wind-farm geometries beyond the aligned and staggered configurations discussed in \cite{ste14g} and to arbitrary wind inflow directions. We start with a short review of the basic concepts of the CWBL model in section \ref{Section_CWBL}. Subsequently we discuss how the CWBL model can be applied to more general configurations and arbitrary wind inflow directions. In sections \ref{section_field} and \ref{section_LES} the CWBL model results are compared with field measurement data for Horns Rev and Nysted \cite{bar09c,bar11}, see figure \ref{figure2}, and LES results for Horns Rev \cite{por13}. The general conclusions are presented in section \ref{Section_discussion_conclusion}.

\section{The coupled wake boundary layer model} \label{Section_CWBL}

The CWBL model used in this work consists of two submodels: (I) a wake model and (II) a ``top-down'' model that are coupled in the following way. Both the Jensen wake \cite{jen83,kat86} and the Calaf {\it et al.} \cite{cal10} ``top-down'' model that are employed in the basic CWBL model assume steady state conditions and have one free parameter each. The models combine the effects of several physical processes using few parameters, e.g.\ in the Jensen model the vertical mixing and wake expansion are represented using the single parameter $\mathrm{k_w}$. While the models perform reasonably well (which is why they continue to be used in practice), in the CWBL approach the free parameter for each model part can be obtained from its complementary parts using an iterative procedure, which will be explained in section \ref{section_coupling}. In this section we start with a short summary of the Jensen wake (section \ref{section_jensen}) and the Calaf {\it et al.} ``top-down'' models (section \ref{sectionTopdown}) used for the basic CWBL approach and then proceed with the introduction of the generalized CWBL two-way coupling procedure (section \ref{section_coupling}).

\subsection{Jensen wake model} \label{section_jensen}
The classic wind-turbine Jensen wake model has been developed based on successive contributions by Lissaman \cite{lis79}, Jensen \cite{jen83}, and Kat\'ic {\it et al.} \cite{kat86}. It assumes that wind-turbine wakes grow linearly with downstream distance and the following expression for the evolution of the wind velocity in the turbine wakes \cite{lis79,jen83}:
\begin{equation}
\label{eq1}
\mathrm{\overline{u}}= \mathrm{\overline{u}_0} \left(1- \frac{1-\sqrt{1-\mathrm{C_T}}}{(1+\mathrm{k_w} x /R)^2} \right) = \mathrm{\overline{u}_0} \left(1- \frac{2a}{(1+\mathrm{k_w} x /R)^2} \right).
\end{equation}
Here $\mathrm{\overline{u}_0}$ is the incoming free stream velocity (the overbar indicates time averaging), $\mathrm{k_w}$ is the wake expansion coefficient, $R$ is the rotor radius, $\mathrm{C_T}=4a(1-a)$ is the thrust coefficient with flow induction factor $a$, and $x$ is the downstream distance with respect to the turbine. Therefore the velocity defect at each point ${\bf x}$ in the wake of a turbine `$j$' located at $(\mathrm{x_j},\mathrm{y_j},\mathrm{z_h})$, where $\mathrm{z_h}$ is the turbine hub-height, is 
\begin{equation}
\label{eq2}
\delta \mathrm{\overline{u}}({\bf x};j) = \mathrm{\overline{u}_0} - \mathrm{\overline{u}} ({\bf x};j) = \frac{2 ~a ~ \mathrm{\overline{u}_0}} {[1+\mathrm{k_w} (x-\mathrm{x_j}) / R]^2}.
\end{equation}
The interaction of the wakes with the ground is modeled by incorporating ``ghost'' or ``image'' turbines under the ground surface \cite{lis79}. Thus, for every turbine $j$ at position $(\mathrm{x_j},\mathrm{y_j},\mathrm{z_h})$ an image turbine at $(\mathrm{x_j},\mathrm{y_j},-\mathrm{z_h})$ is added. Wake superposition effects are accounted for by adding the squared velocity deficits of interacting wakes at the point ${\bf x}$ to obtain
\begin{equation}
\mathrm{\overline{u}} ({\bf x}) = \mathrm{\overline{u}_0} - \sqrt{ \sum_{j} [\delta \mathrm{\overline{u}} ({\bf x};j)]^2 },
\label{eq-velsuperposed}
\end{equation}
where $\delta \mathrm{\overline{u}} ({\bf x};j)$ is the velocity deficit that would exist at the point ${\bf x}$ if only a single turbine Ò$j$Ó upstream were to cause a single wake, while $\mathrm{\overline{u}} ({\bf x})$ refers to the velocity at a given location, and the summation $j$ is over all turbine (and ghost turbine) wakes. The predicted turbine power ${\mathrm P_\mathrm{T}}$ of turbine `$\mathrm{T}$' (where the subscript $\mathrm{T}$ refers to a particular turbine and ${\bf x}_{T,k}$ are all the $\mathrm{N_d}$ discrete positions, i.e.\ the grid points used in the numerical evaluation, comprising the turbine `$\mathrm{T}$') is given by
\begin{equation}
\frac{\mathrm{P_T}}{\mathrm{P_1}} = \left(\frac{1}{\mathrm{N_d}}\sum_{k=1}^{\mathrm{N_d} } \frac{ \mathrm{\overline{u}} ({\bf x}_\mathrm{T,k})}{ \mathrm{\overline{u}_0}}\right)^3. 
\end{equation}
The summation over $k$ is over all points on the turbine disk and the velocity ratios are obtained from equation \eqref{eq-velsuperposed}. The resulting power ratio represents the power normalized with the power of a free-standing turbine $\mathrm{P_1}$ (or that of turbines in the first row of the wind-farm). Following Lissaman \cite{lis79} it is common to define the wake decay parameter as $\mathrm{k_w}=\kappa /\ln(\mathrm{z_h}/\mathrm{z_{0,lo}})$, where $\mathrm{z_{0,lo}}$ is the roughness length of the ground surface, and $\kappa=0.4$ is the von K\'arm\'an constant. 

\subsection{The ``top-down'' model} \label{sectionTopdown}
The ``top-down'' wind-farm model also traces its origins to Lissaman \cite{lis79}. It was further developed and presented in updated forms by Frandsen \cite{fra92,fra06} and Calaf {\it et al.} \cite{cal10}. This model is essentially a single column model of the ABL based on momentum theory. The objective of the ``top-down'' model is to predict the horizontally ($\langle . \rangle$) and time ($\overline{{\color{white}a}}$) averaged velocity profile $\langle \mathrm{\overline{u}} \rangle(z)$ in the wind-turbine array boundary layer that is assumed to be spatially fully developed so that averaging in both the streamwise and spanwise directions makes sense. If there is no wind-farm the flow can be assumed to be undisturbed, and we have a velocity profile that follows the traditional logarithmic law:
\begin{align}
\label{Eq_profile_5}
 \langle \mathrm{\overline{u}_0} \rangle (z) = \frac{u_{*}}{\kappa} \ln \left( \frac{z}{\mathrm{z_{0,lo}}} \right) & ~~~~~~~~~~ \mbox{for} & \mathrm{z_{0,lo}} & \leq z \leq \delta_\mathrm{H},
\end{align}
where $u_*$ is the friction velocity and $\delta_\mathrm{H}$ indicates the height of the ABL. The ``top-down'' model approach from Calaf {\it et al.} \cite{cal10} assumes the presence of two constant stress layers, one above and one below the turbine hub-height, and a wake layer in between. This model can be used to obtain the ratio of the mean velocity at hub-height in the fully developed region of the wind-farm to the incoming reference velocity as
\begin{equation}
\label{Eq_veloc}
 \frac{\langle \mathrm{\overline{u}} \rangle(\mathrm{z_h}) }{\langle \mathrm{\overline{u}_0} \rangle(\mathrm{z_h})} = \frac{\ln \left(\delta_\mathrm{H}/{\mathrm{z_{0,lo}}} \right)} {\ln \left( \delta_\mathrm{H}/{\mathrm{z_{0,hi}}} \right)} ~\ln \left[ \left( \frac{\mathrm{z_h}}{{\mathrm{z_{0,hi}}}} \right) \left(1 + \frac{D}{2\mathrm{z_h}} \right)^\beta \right] \left[ \ln \left( \frac{\mathrm{z_h}}{{\mathrm{z_{0,lo}}} } \right)\right]^{-1},
\end{equation}
where $\mathrm{z_{0,hi}}$ denotes the roughness length of the wind-farm, which is defined as
\begin{equation} \label{Eq_defz0hi}
\mathrm {z_{0,hi}}= \mathrm{z_h} \left(1+\frac{D}{2 \mathrm{z_h}}\right)^{\beta} \\
\exp \left(- \left[ \frac{\pi \mathrm{C_T}}{8 \mathrm{w_f}\mathrm{s_x}\mathrm{s_y} \kappa^2} + \left( \ln \left[ \frac{\mathrm{z_h}}{{\mathrm { z_{0,lo}}}} \left( 1 -\frac{D}{2\mathrm{z_h}}\right)^{\beta}	\right] \right)^{-2}		\right]^{-1/2} 				\right),
\end{equation}
with $\beta=\nu_w^*/(1+\nu_w^*)$ and $\nu_w^*\approx 28 \sqrt{ \pi \mathrm{C_T}/ (8 \mathrm{w_f}\mathrm{s_x}\mathrm{s_y} )}$. Here, $\mathrm{w_f}$ indicates the effective wake area coverage, which will be further discussed in section \ref{section_coupling} and the appendix (note that the product $\mathrm{w_f}\mathrm{s_x}\mathrm{s_y}$ was denoted as $\mathrm{s_x} \mathrm{s_{ye}}$ in Ref.\ \cite{ste14g}, where $\mathrm{s_{ye}}$ was defined as the effective spanwise distance). The power output ratio is defined as the ratio of the power output of turbines in the fully developed region to that of the turbines in the first row, i.e.\ 
\begin{equation}
\label{Eq_Power}
\frac{\mathrm{P_\infty}}{\mathrm{P_1}}=\left( \frac{\langle \mathrm{\overline{u}} \rangle(\mathrm{z_h}) }{\langle \mathrm{\overline{u}_0} \rangle(\mathrm{z_h})} \right)^3.
\end{equation}
\subsection{Generalized CWBL approach} \label{section_coupling}

\begin{figure}
\centering
\subfigure{\includegraphics[width=0.999\textwidth]{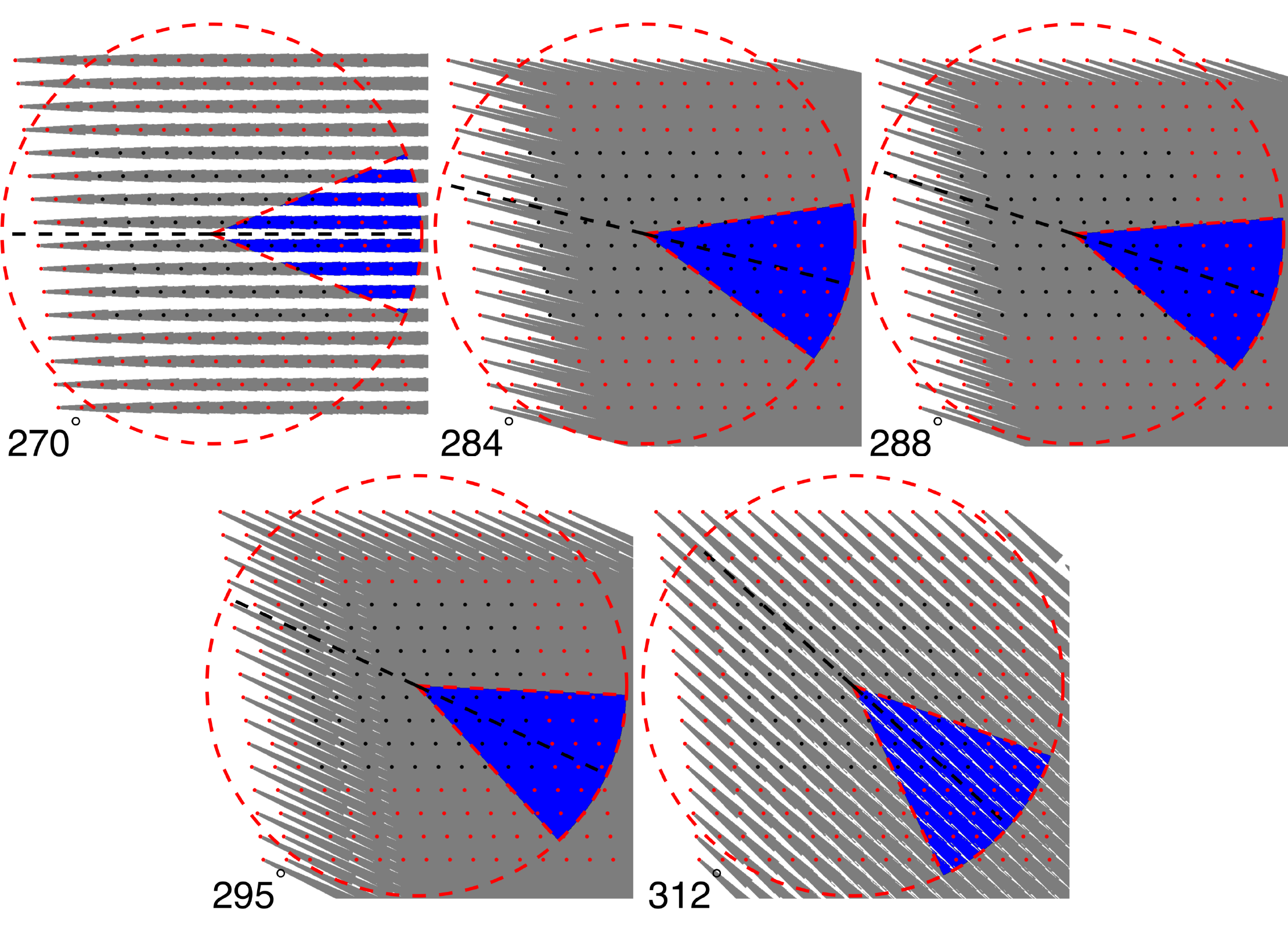}}\\
\caption{An ``extended" $16\times16$ Horns Rev wind-farm used to illustrate the iterative procedure, see section \ref{section_coupling} for details. The black dots indicate the turbine positions in the actual wind-farm and the red dots denote the added turbines that are used to create the ``extended" farm that ensures a fully developed region is reached for the purpose of computing the effective wake area $\mathrm{w_f}$. The red dashed circle has a diameter $\mathrm{D_{wf}} = \sqrt{(4 \mathrm{A_{wf}}/ \pi)}$, where $\mathrm{A_{wf}}$ is the total area of wind-farm, and the center of the circle located at the center of mass of the wind-turbines. The $45^\circ$ pie slice of the circle indicates the sector over which the effective wake area coverage $\mathrm{w_f}$ is determined, which is centered around the wind direction under consideration (dashed black line). The gray areas indicate the wake areas using the criterion $\mathrm{\overline{u}(x,y)} < 0.95 \mathrm{\overline{u}_0}$. In the pie sector the wake area is colored blue. The effective wake area coverage $\mathrm{w_f}$ is determined using equation \eqref{equation_wf}, which is equivalent to dividing the blue area by the area of the pie sector. The plots for $270^\circ$ ($\mathrm{w_f}=0.56$), $284^\circ$ ($\mathrm{w_f}=1$), $288^\circ$ ($\mathrm{w_f}=1$), $295^\circ$ ($\mathrm{w_f}=1$), and $312^\circ$ ($\mathrm{w_f}=0.90$) correspond to the results shown in figure \ref{figure6}.}
\label{figure0}
\end{figure}

The two-way coupling between the wake and ``top-down" modules of the CWBL model is obtained in the fully developed region of the wind-farm and is enforced through an iterative procedure. In the present implementation of the CWBL model we assume that the fully developed region is reached at the $10^\mathrm{th}$ turbine row, measured in the downstream wind direction, although other criteria can be used to define this region. In practice, the $10^\mathrm{th}$ row criterion ensures that the Jensen model has reached fully developed conditions for all of the cases we have considered. The iteration procedure determines the effective wake area coverage $\mathrm{w_f}$ in equation \eqref{Eq_defz0hi} using the wake (Jensen) model and obtains the wake expansion coefficient $\mathrm{k_{w,\infty}}$ by matching the velocity at hub-height in the Jensen and ``top-down" models. Here $\mathrm{k_{w,\infty}}$ indicates the wake expansion rate in equations \eqref{eq1} and \eqref{eq2} for the fully developed region of the wind-farm. Note that the effective wake area coverage $\mathrm{w_f}$ is used to calculate the interaction between the wind and the ABL in the ``top-down" model portion of the CWBL model. In the traditional implementation of the ``top-down" model a horizontal area equal to $(\mathrm{s_x} D)(\mathrm{s_y} D)$ across which the vertical fluxes per turbine are evaluated is considered. However, in the CWBL model, the control volume that is used in the ``top-down" model part is smaller and equal to $(\mathrm{s_x} D)(\mathrm{w_f} \mathrm{s_y} D)$, where the factor $\mathrm{w_f}$ ($\mathrm{w_f} \leq 1$) is the fraction of the wind-farm area in which wakes affect the mean velocity appreciably (a more precise definition will be provided below).

In the Jensen model the mean velocity in the fully developed region depends on $\mathrm{k_{w,\infty}}$, while in the ``top-down" model this mean velocity depends on the effective wake area coverage $\mathrm{w_f}$. Therefore the effective wake area coverage $\mathrm{w_f}$ and the wake expansion rate $\mathrm{k_{w,\infty}}$ need to be iterated until the mean velocity in the fully developed region predicted using both models converge to within $0.1\%$. In the appendix we describe how this approach relates to the original CWBL iteration procedure introduced in Ref.\ \cite{ste14g} that was straightforward to apply only to the cases of aligned and staggered wind-farms.

We now detail the iteration procedure for each wind direction. \\

\noindent Step 1: Assume an initial value for the wake expansion parameter in the fully developed region $\mathrm{k_{w,\infty}}$. For the first iteration assume $\mathrm{k_{w,\infty}}=\mathrm{k_{w,0}}=\kappa /\ln(\mathrm{z_h}/\mathrm{z_{0,lo}})$, where $\mathrm{k_{w,0}}$ is the wake expansion coefficient at the entrance of the wind-farm that can be computed based on $z_{\mathrm h}$ and known values of $\mathrm{z_{0,lo}}$.
In the iterative procedure we use $\mathrm{k_{w,\infty}}$ for all of the turbines in the wind-farm to approximate the fully developed region more closely in the wake (Jensen) model. \\
\\ 
\noindent Step 2: Use the (Jensen) wake model to calculate the effective wake area coverage $\mathrm{A_{wake}}$ in the fully developed region. For wind-farms that are not very large, such as Horns Rev and Nysted, shown in figure \ref{figure2}, we first extend the wind-farms by replication of the turbine array in order to obtain converged (Jensen) wake model results for all wind directions. For example, in the present applications to the Horns Rev and Nysted wind-farms, we extend the wind-farm array to a $16\times16$ array of turbines, which has been selected to make sure that $\mathrm{w_f}$ is converged for all wind directions, while it is still sufficiently small to allow efficient calculations.
 Subsequently we define a $45$ degree pie slice of a circle. The circle has an equivalent wind-farm diameter $\mathrm{D_{wf}} = \sqrt{(4 \mathrm{A_{wf}}/ \pi)}$, where $\mathrm{A_{wf}}$ is the total area of the extended wind-farm. The center of the circle is located at the center of mass of the extended wind-farm. The center-line of the pie slice is aligned with the wind direction, see figure \ref{figure0}. Subsequently we calculate the wake area inside the pie-shaped region, $\mathrm{A_{wpr}}$, as the area where $\mathrm{\mathrm{\overline{u}}(x,y)} < 0.95 \mathrm{\overline{u}_0}$ (blue area in figure \ref{figure0}). The effective wake area coverage $\mathrm{w_f}$ is then defined as 
\begin{equation} \label{equation_wf}
\mathrm{w_f}= \frac{\mathrm{A_{wpr}}}{\mathrm{A_{pr}}},
\end{equation}
where $\mathrm{A_{pr}}=(1/8)\mathrm{A_{wf}}$ is the entire area of the $45$ degree pie region of the circle with diameter $\mathrm{D_{wf}}$. A $45$ degree angle ensures that the area over which $\mathrm{w_f}$ is computed is large enough to obtain converged results for $\mathrm{w_f}$, but small enough to appropriately represent wind-farm characteristics in a particular direction. Although the exact value, $45$ degrees, is somewhat arbitrary, the results are robust with respect to the precise value of the angle, as long as the area that is used to evaluate $\mathrm{w_f}$ is sufficiently large. Therefore we took a constant angle in our calculations for conceptual simplicity. The $5\%$ velocity threshold that is used to determine the wake area has been determined empirically to give results similar to those obtained using the original CWBL method (see the appendix) and could be different when another wake model is used in the CWBL framework.

It should be noted that the wake region that is defined in Step 2 is not the same as what is usually considered the turbine wake. The latter, in the case of the Jensen model, is equal to an ever expanding ``cone" in which the velocity defect decreases indefinitely. Instead, based on the velocity threshold introduced in the current CWBL approach, for the case of isolated turbines, the velocity deficit will fall below the threshold at some distance downstream of each turbine. This point will then be ``outside" the wake again, i.e.\ the wake will ``terminate". For a wind-farm, the velocity defect superposition affects where the threshold is reached also in the lateral direction. For example, in the $270^\circ$ case in figure \ref{figure0}a, the geometric series convergence properties of the superposition leads to the ``wake area region" ceasing to grow in the cross-stream direction at a certain downstream position (as can be seen in the figure, the grey area no longer grows downstream, and this is because of the velocity threshold). This behavior is consistent with Frandsen's view \cite{fra06} of a vertical-only process since the absence of further horizontal growth implies that the only fluxes occur in the vertical direction.\\
\\
\noindent Step 3: Calculate $\langle \mathrm{\overline{u}} \rangle(\mathrm{z_h}) / \langle \mathrm{\overline{u}_0} \rangle(\mathrm{z_h})$ using equation \eqref{Eq_veloc} with the ``top-down'' model and the effective wake area coverage fraction $\mathrm{w_f} $ evaluated in step 2 to find the power production in the fully developed region.\\

Iterate between steps 2 and 3 until the velocity at hub-height found by the ``top-down" and wake (Jensen) model for the fully developed region agree to within $0.1\%$.\\
 
\noindent In order to model the entrance effects, i.e.\ the development of the average power production of the turbines in the first several rows, the CWBL uses an interpolation between $\mathrm{k_{w,0}}$ and $\mathrm{k_{w,\infty}}$ to assign the wake coefficient $\mathrm{k_{w,T}}$ for each turbine in the wind-farm using
\begin{equation}
\label{equation_finitewindfarm}
\mathrm{k_{w,T}}= \mathrm{k_{w,\infty}}+(\mathrm{k_{w,0}}-\mathrm{k_{w,\infty}}) \exp(-\zeta m),
\end{equation}
where $m$ is the number of turbine wakes (not including ghost wakes) that overlap with the turbine of interest and $\zeta = 1$. As indicated in Ref.\ \cite{ste14g} this value of $\zeta$ is an empirical choice (based on using the Jensen model), which ensures that there is a relatively sharp transition from the entrance region to the fully developed region, which is consistent with the behavior observed in the data. Note that this approach means that the wake (Jensen) part of the CWBL model dominates in the entrance region of the wind-farm, while the wake development further downstream is primarily determined by the ``top-down'' model. 

\begin{figure}
\centering
\subfigure{\includegraphics[width=0.49\textwidth]{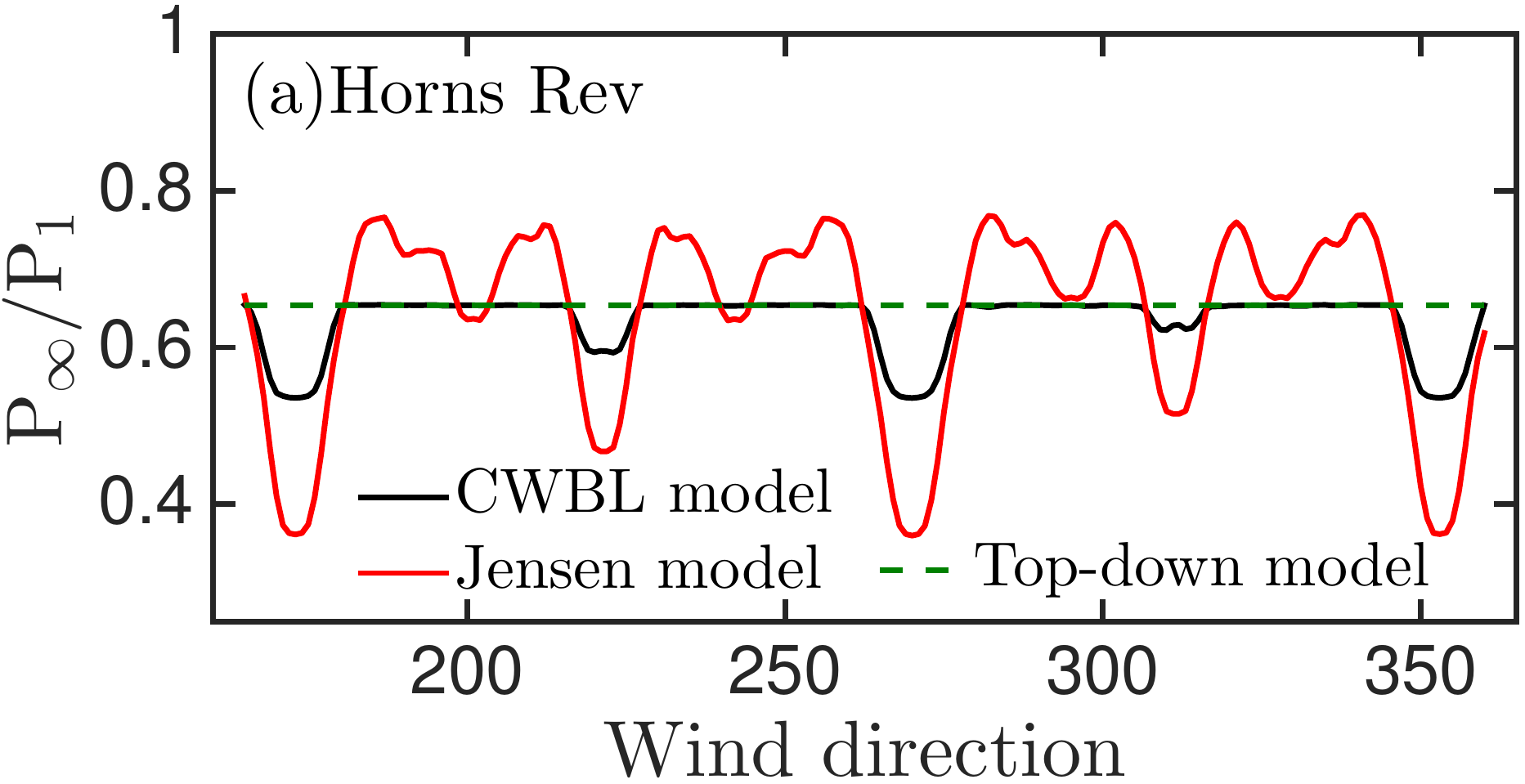}}
\subfigure{\includegraphics[width=0.49\textwidth]{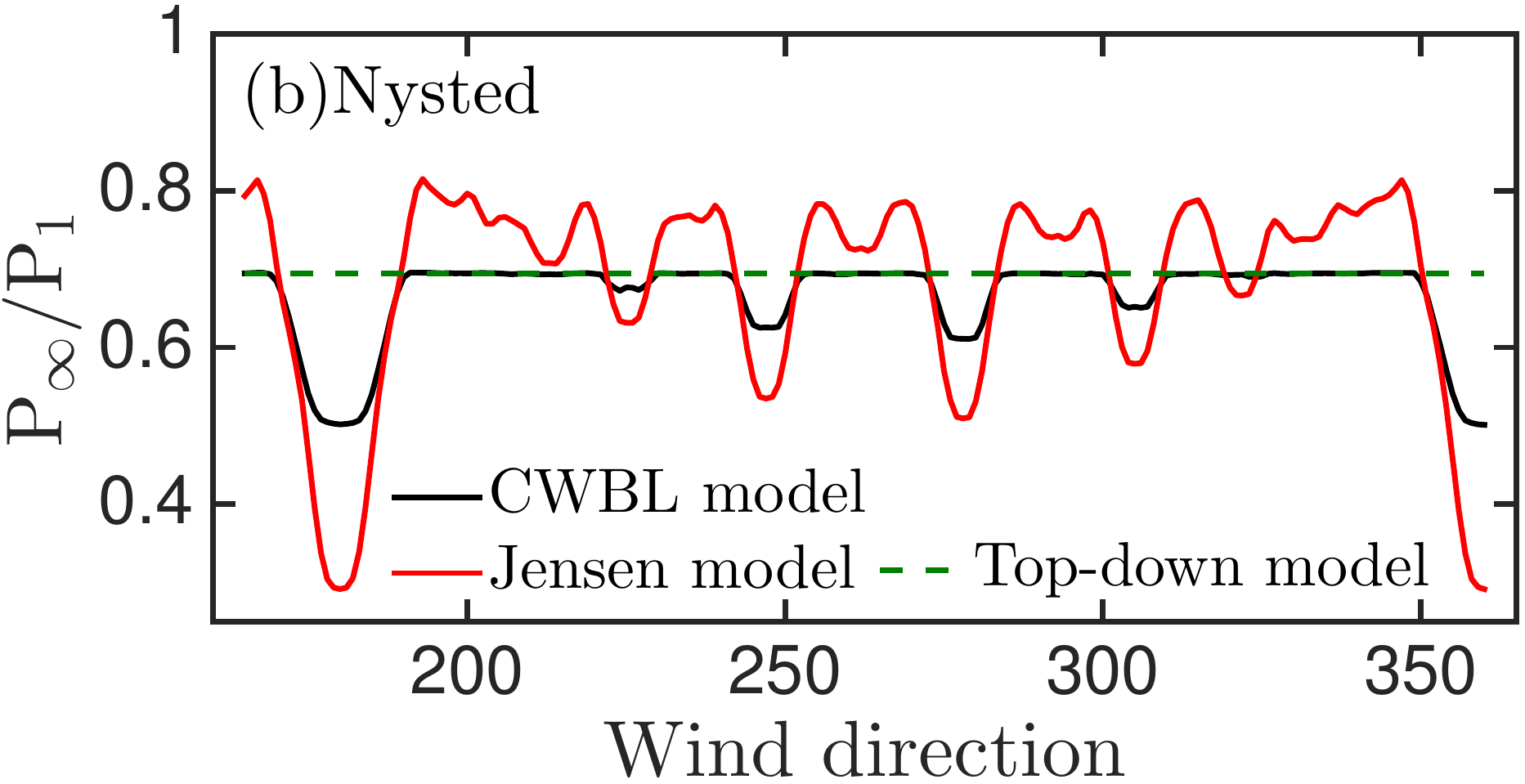}}
\caption{Power output ratio $\mathrm{P_\infty}/\mathrm{P_1}$ for (a) Horns Rev and (b) Nysted obtained from the CWBL and Jensen model. Here $\mathrm{P_\infty}$ is the power output of the turbine at the end of the extended $16\times16$ Horns Rev / Nysted wind-farms, used in the iteration procedure; see section \ref{section_coupling} for details. Note that the CWBL model predicts stronger wake effects for the non-aligned wind directions. A comparison with the field measurements in figure \ref{figure4} shows that the stronger wake effects predicted by the CWBL model for the non-aligned wind directions show better agreement with the field data. The dashed horizontal lines indicate the ``top-down" model predictions for the fully developed region.}
\label{figure5}
\end{figure}

\begin{figure}
\centering
\subfigure{\includegraphics[height=0.24\textwidth]{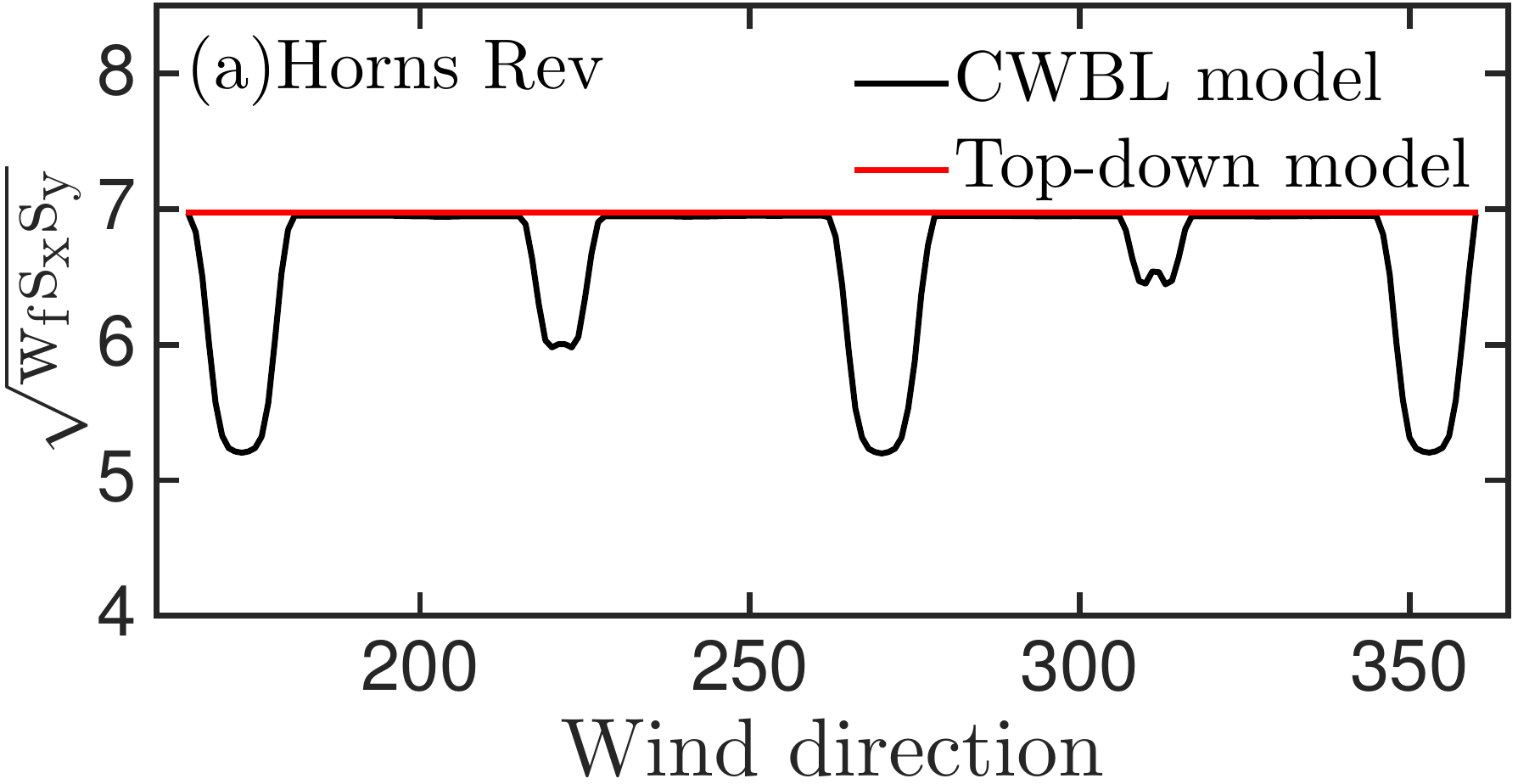}}
\subfigure{\includegraphics[height=0.24\textwidth]{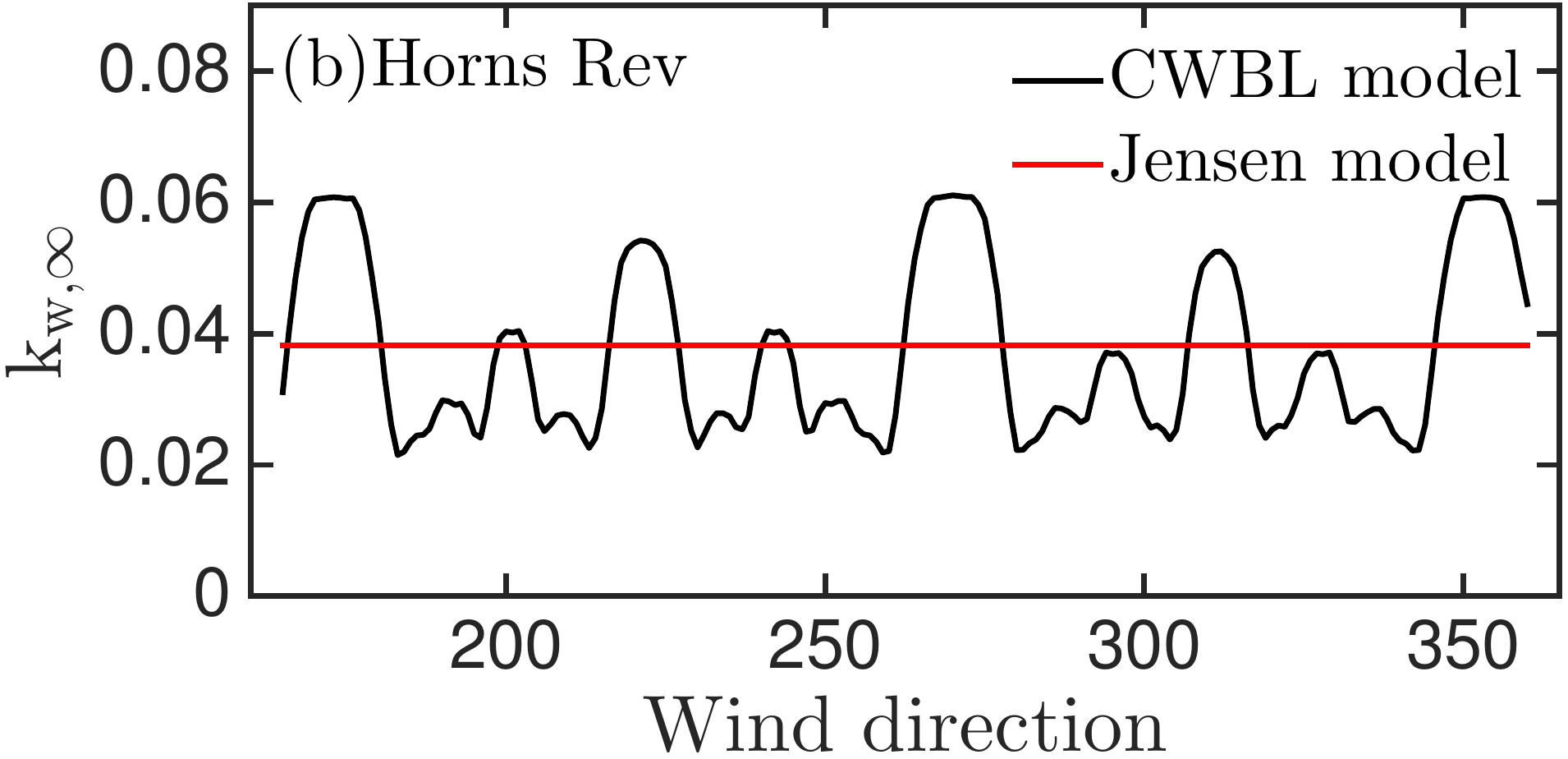}}
\subfigure{\includegraphics[height=0.24\textwidth]{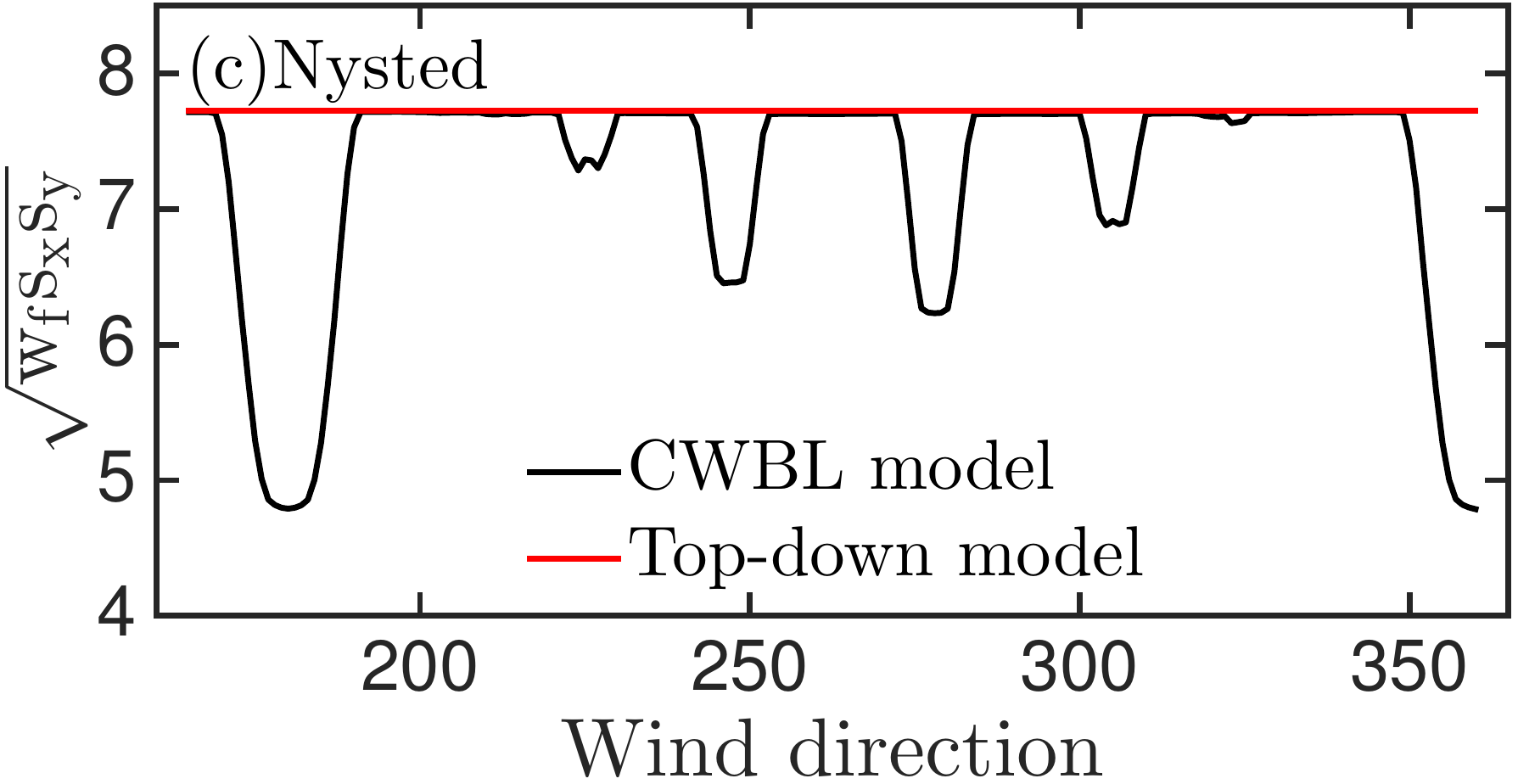}}
\subfigure{\includegraphics[height=0.24\textwidth]{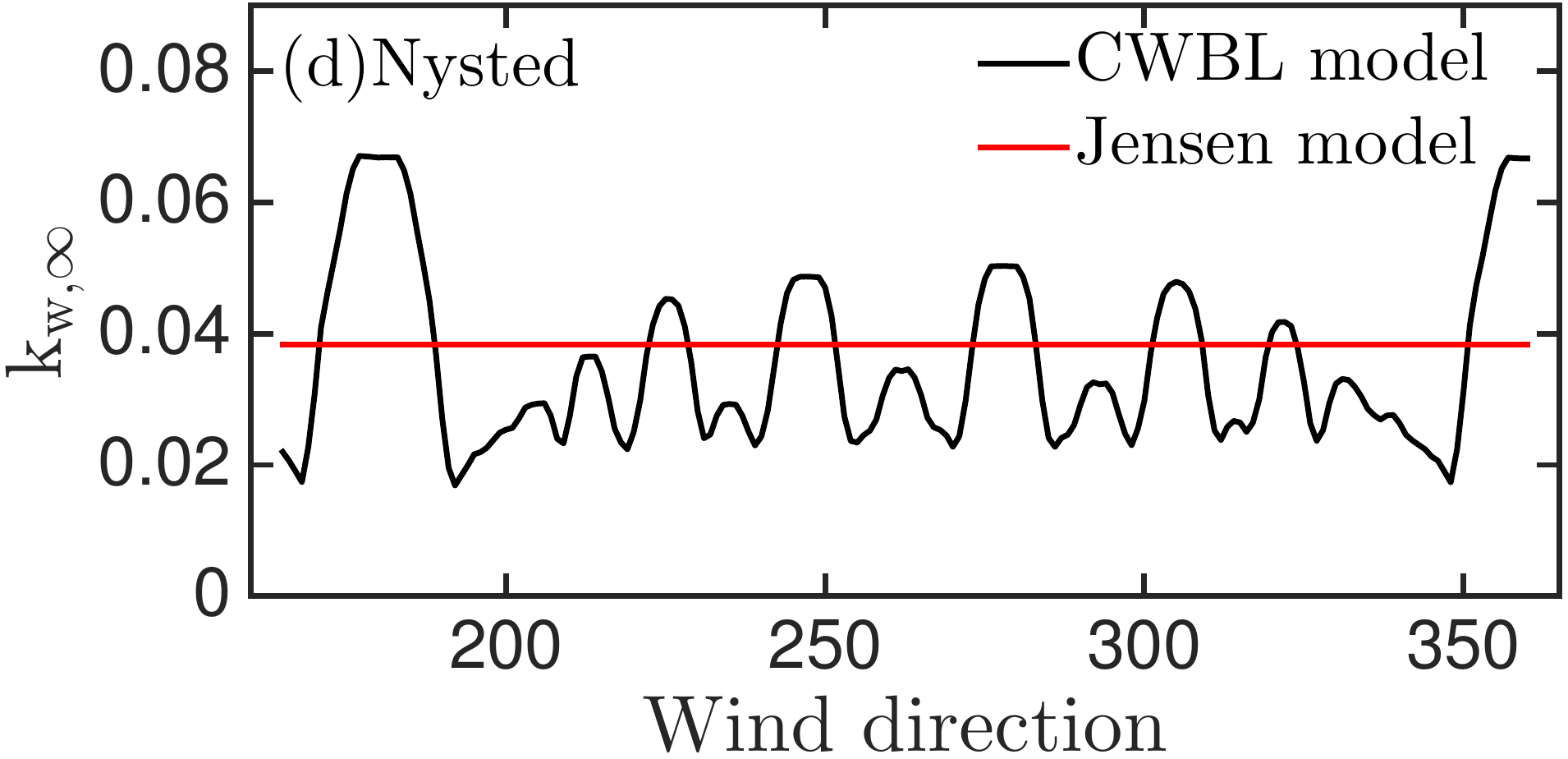}}
\caption{(a,c) The black line indicates $\sqrt{\mathrm{w_f}\mathrm{s_x}\mathrm{s_y}} $ in the CWBL model and the red line $\mathrm{s_x}\mathrm{s_y}$ in the ``top-down" model as function of the wind direction for Horns Rev and Nysted. (b,d) The black line indicates $\mathrm{k_{w,\infty}}$ in the CWBL model as function of the wind direction and the red line denotes the value of $\mathrm{k_{w,0}}$ used in the CWBL and Jensen model results shown in figures \ref{figure5} and \ref{figure4}.}
\label{figure3}
\end{figure}

We note that the generalized CWBL model used here assumes that all turbines operate in region II, in which the thrust coefficient $\mathrm{C_T}$ can be considered to be approximately constant as function of the wind speed \cite{joh04}. Note that the LES data \cite{por13} and field measurements \cite{bar09c,bar11} to which we compare our results are obtained for a wind-speed of $\approx8\pm0.5$m/s, which corresponds to the turbines operating in region II. In addition the model neglects the variation of the power coefficient $\mathrm{C_P}$ with wind-speed. For the comparisons discussed here the data have been obtained for a very narrow range of wind speeds and hence it is indeed reasonable to assume that $\mathrm{C_P}$ is the same for the front turbine as for the downstream ones (i.e.\ we are assuming that each turbine adjusts its tip-speed ratio so as to operate at the optimal $\mathrm{C_P}$). These assumptions imply that the ratio of power is the same as the ratio of cubed velocities; we use this relationship to obtain mean power predictions from the velocities computed based on the model. The CWBL model can also use different $\mathrm{C_T}$ values for turbines in the entrance region of the wind-farm but maintain a constant $\mathrm{C_T}$ value in the fully developed region where the coupling between the wake and ``top-down" part of the model is applied \cite{ste14g}.

We note that, in contrast to the $\mathrm{s_{ye}}$ based CWBL model \cite{ste14g}, where we performed the calculations of the Jensen model in three dimensions, we now perform these calculations in a two dimensional plane at hub-height, but still include the image (ghost) turbine wakes. We have verified that this gives very similar results to the 3D approach. The only difference is the representation of partial wake effects in the vertical direction, which are not accounted for in the two dimensional model calculations.

Figure \ref{figure5} shows the power output ratio for (a) Horns Rev and (b) Nysted obtained from the CWBL and Jensen model. Results indicate that for non-aligned wind directions the CWBL model predicts a lower turbine power output in the fully developed region of the wind-farm than the Jensen model. Figures \ref{figure3}a and \ref{figure3}c show the effective geometric mean turbine spacing $\sqrt{\mathrm{w_f}\mathrm{s_x}\mathrm{s_y}}$ as function of the wind direction determined using the CWBL model for Horns Rev and Nysted. The figure shows that $\sqrt{\mathrm{w_f}\mathrm{s_x}\mathrm{s_y}}$ is smaller than the actual geometric mean turbine spacing $s=\sqrt{\mathrm{s_x}\mathrm{s_y}}$ for wind directions that correspond to there being a relatively small distance between consecutive downstream turbine rows. Figure \ref{figure3}b and \ref{figure3}d show the corresponding $\mathrm{k_{w,\infty}}$ values determined with the CWBL model. In agreement with earlier results \cite{ste14g}, we find that the wake decay exponent predicted by the CWBL model is higher for wind directions aligned with the wind-farm, and when the downstream distance between the turbines is smaller. This higher wake expansion coefficient captures the effect of the added wake turbulence for wind directions aligned with the wind-farm layout. 

The reason for the correlation of the wake expansion coefficient with turbulence levels is that for wind-farm configurations which ``load" the ABL more, e.g. that have smaller inter turbine spacings, the ``top-down" model predicts a higher friction velocity above the wind-farm ($\mathrm{u_{*,hi}}$). This predicted higher friction velocity implies that the turbulence levels should be higher due to the  correlation between velocity variance and shear stress (momentum flux) typically expected in turbulent boundary layers. To obtain the matching with the Jensen-model predicted velocity in the ``deep-array" portion, the CWBL model then adjusts the $\mathrm{k_{w,\infty}}$ value, which reproduces the correlation between $\mathrm{k_{w,\infty}}$ and higher turbulence intensities. Note, however, that for non-aligned directions the wake expansion factor predicted by the model can be smaller than the wake expansion coefficient that is predicted for turbines in the first row. This trend is perhaps surprising since one would expect that under any conditions the turbulence level inside the wind-farm should exceed (or at least be equal to) that of the inflow. While this trend is not easy to explain, we note that it arises from matching the velocities in the far field between two models (Jensen and ``top-down") which each include a number of approximations. Also, we note that the friction velocity below the turbine height ($\mathrm{u_{*,lo}}$) in the ``top-down" model is in fact predicted to be smaller than the incoming $\mathrm{u_*}$ due to the reduced shear under the turbines. Therefore, it is possible that some reduced momentum exchanges may occur in the presence of wind-farms, under certain conditions. \\

\section{Comparison with field measurements} \label{section_field}

\begin{table}
\caption{Model parameters employed for the calculations of the CWBL and the Jensen model. The field measurement \cite{bar09c,bar11} and LES \cite{por13} data to which we compare are for an average wind speed of $8 \pm 0.5$ m/s.}
\label{table1}
\begin{center}
\vspace{-10pt}
\begin{tabular}{| l |c|c|c|c|c|c|c|}
\hline
Model parameters	 									&	Horns Rev 	& Nysted			\\ \hline	
Aligned configuration 									&	270$^\circ$		& 278$^\circ$ 		\\ \hline	
Streamwise distance aligned ($\mathrm{s_x}$)					&	7.00				& 10.40 			\\ \hline	
Spanwise distance aligned ($\mathrm{s_y}$)					&	6.95				& 5.74 			\\ \hline	
Hub-height	 ($\mathrm{z_h}$)							&	70 m	 			& 69 m 			\\ \hline	
Turbine diameter ($D$)									&	80 m				& 82.4 m 			\\ \hline	
Thrust coefficient ($\mathrm{C_T}$)							&	0.78	 			& 0.78 			\\ \hline	
Turbulence Intensity										& 	7.7$\%$			& 7.7$\%$			\\ \hline
Roughness length ground ($\mathrm{z_{0,lo}}$)				&	0.002 m			& 0.002 m			\\ \hline
Maximum internal boundary layer height ($\delta_\mathrm{H}$)				&	500 m			& 500 m			\\ \hline
(Initial) Wake expansion coefficient $\mathrm{k_{w,0}}$			&	0.0382			& 0.0382			\\ \hline

\end{tabular}
\end{center}
\end{table}

\begin{figure}
\centering
\subfigure{\includegraphics[width=0.99\textwidth]{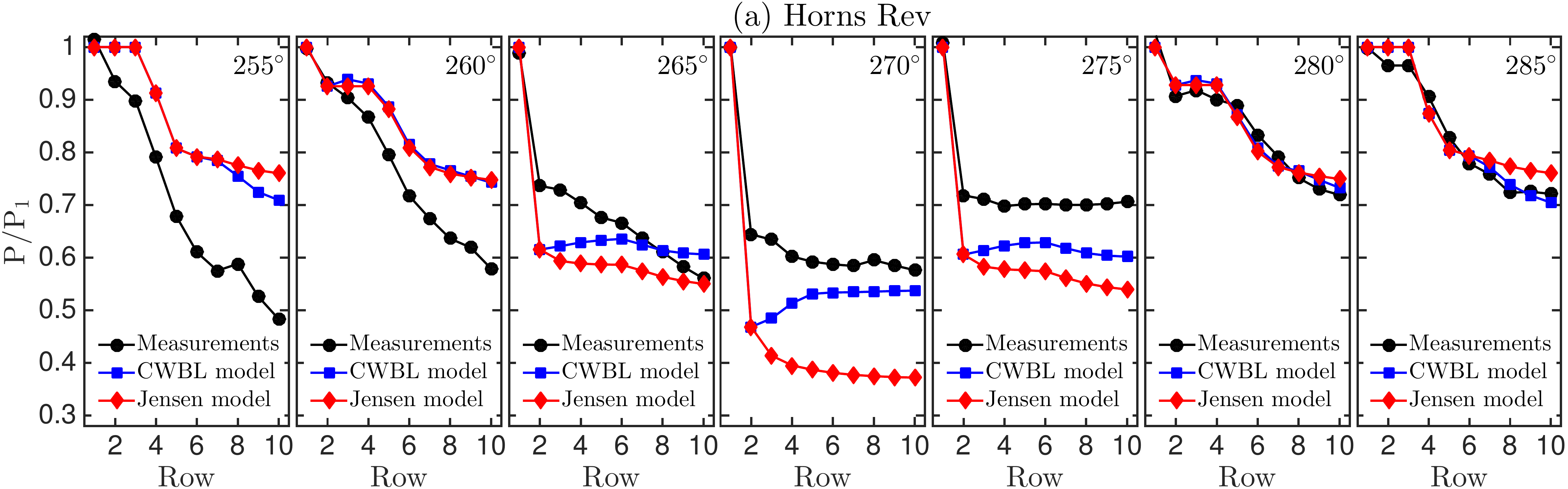}}
\subfigure{\includegraphics[width=0.99\textwidth]{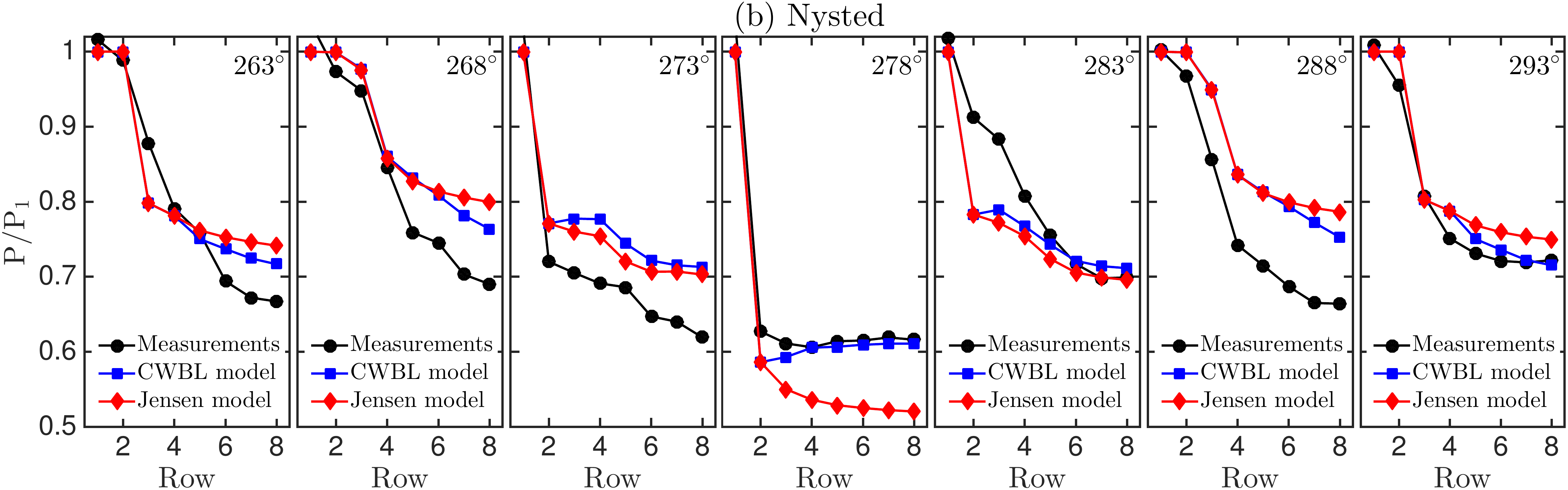}}
\caption{Comparisons between field measurements (black circles) for Horns Rev (a, top panel) and Nysted (b, lower panel) \cite{bar09c,bar11} with the CWBL model (blue squares) and the Jensen model (red diamonds). The field measurement results are reported for a 5 degree wind sector. The models were run with $0.5$ degree increments to match the $5$ degree wind sector data from the field experiments, e.g.\ for $270^\circ$ the results is the average over 11 cases between $267.5^\circ-272.5^\circ$. The field measurement data by Barthelmie {\it et al.} \cite{bar09c,bar11} are digitally extracted from their figures.}
\label{figure4}
\end{figure}

\begin{figure}
\centering
\subfigure{\includegraphics[width=0.99\textwidth]{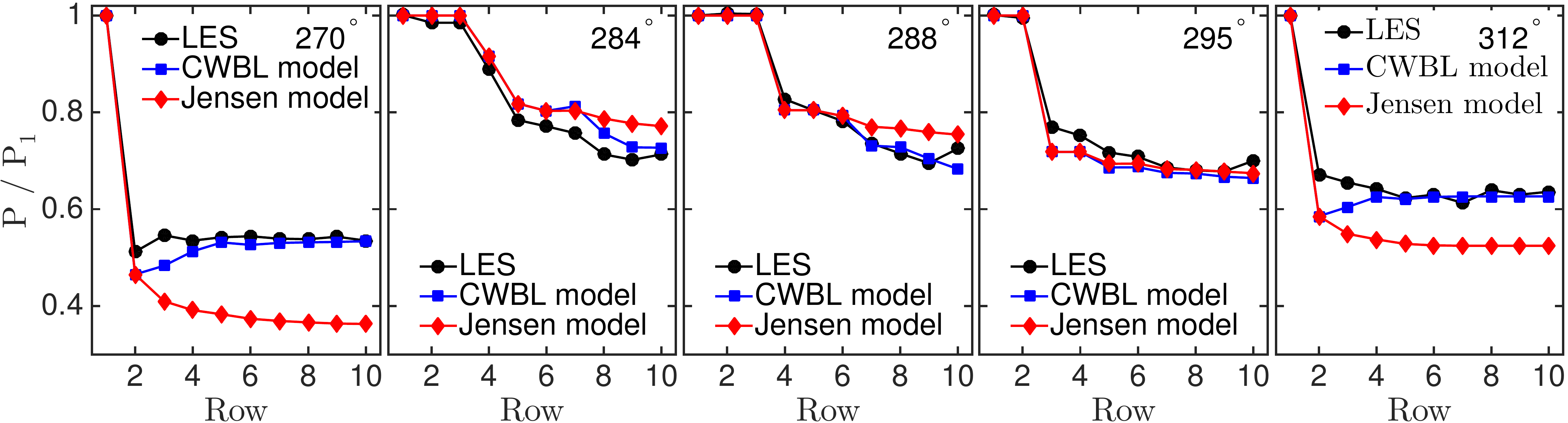}}\\
\subfigure{\includegraphics[width=0.99\textwidth]{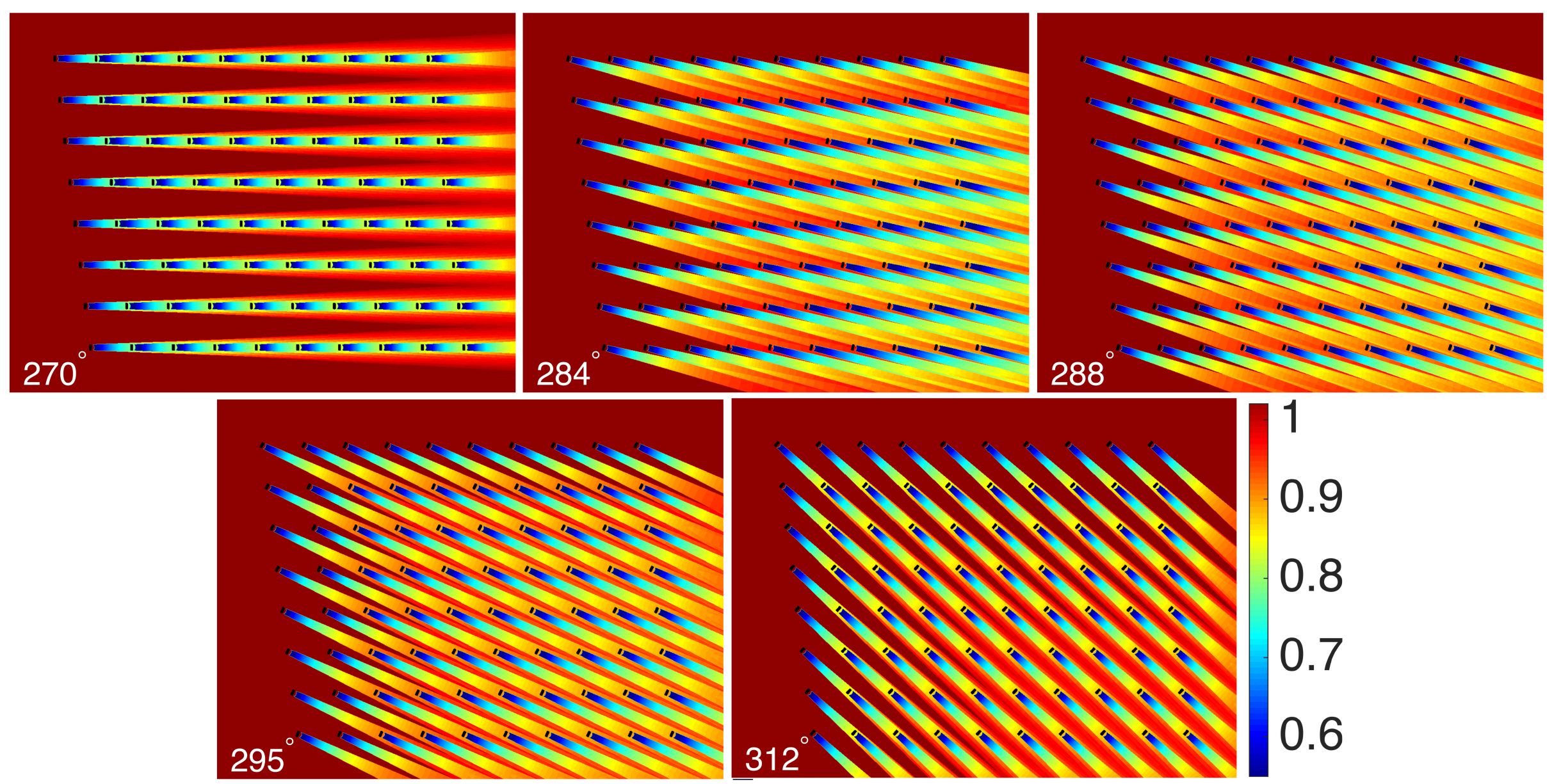}}\\
\caption{Upper panels: Comparison between the LES results from Port\'e-Agel {\it et al.} \cite{por13} (black circles), the CWBL model (blue squares), and the Jensen model (red diamonds) for the power production in the Horns Rev wind-farm. The top panels indicate the results for the wind directions $270^\circ$, $284^\circ$, $288^\circ$, $295^\circ$, and $312^\circ$. Both the LES and model results are averaged over columns C2 to C4, see figure \ref{figure2}. The LES data from Port\'e-Agel {\it et al.} \cite{por13} are digitally extracted from their figures. Lower panels: Visualization of the normalized velocity field at hub-height obtained from the CWBL model for these cases. The color indicates the velocity normalized with the incoming hub-height velocity.}
\label{figure6}
\end{figure}

\begin{figure}
\centering
\subfigure{\includegraphics[width=0.99\textwidth]{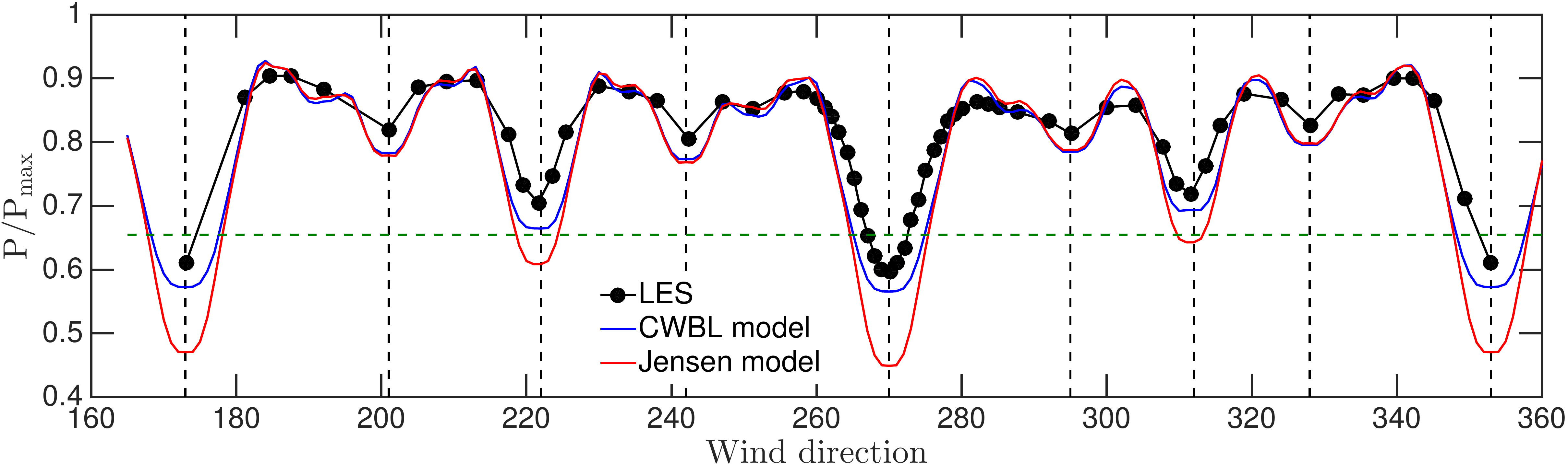}}
\caption{Normalized total power output $\mathrm{P} / \mathrm{P_{max}}$ of Horns Rev as function of the incoming wind direction, where $\mathrm{P_{max}}$ is the power of a non wake affected turbine times the number of turbines in Horns Rev. The LES results by Port\'e-Agel {\it et al.} \cite{por13} are indicted by black circles, which are digitally extracted from their figures. The CWBL and Jensen model are denoted by blue and red lines, respectively. The dashed horizontal line gives the ``top-down" model's prediction for Horns Rev wind-farm. Note that the CWBL model captures the power output for the non-aligned wind inflow directions of $173^\circ$, $222^\circ$, $270^\circ$, $312^\circ$, and $353^\circ$ more accurately than the Jensen model. Evaluating the rms of the mean power predictions to LES data, i.e.\ $\mathrm{\sqrt{ \langle [(P_{pred}(\phi) -P_{LES}(\phi))/ P_{LES}(\phi)]^2 \rangle}}$ over all wind directions $\phi$, we find that the CWBL results differ $6.3\%$ from the LES results while the Jensen model results differ $9.5\%$ from the LES results. This gives an indication of the improvement that can be expected, however it should be noted that the observed improvement will depend on the case that is considered.}
\label{figure7}
\end{figure}

Figure \ref{figure4} compares the CWBL and Jensen model predictions with field measurement data from Horns Rev and Nysted \cite{bar09c,bar11}. According to the Upwind report \cite{bar11} (see chapter 2, page 38 and 42 for the Horns Rev case) the data from the field measurements have been analyzed such that edge effects of the wind-farm are not included. Thus, for each row only turbines which are actually experiencing wake effects are included in the analysis. Therefore we show in figure \ref{figure4} the power output of turbines in columns C5-C7 (C6-C8) for the wind directions $255^\circ - 265^\circ$ ($263^\circ - 273^\circ$) and from columns C2-C4 for the wind directions $270^\circ - 285^\circ$ ($278^\circ - 293^\circ$) for Horns Rev (Nysted). According to Barthelmie {\it et al.} \cite{bar10b} the power in each column is normalized by its own upstream leading turbine, and then the ``average normalized power'' is evaluated and plotted as a function of distance. However, note that since in the model the power of the first turbine is always exactly the same, such distinctions (normalize first and then average or average first and then normalize) do not affect the model predictions. Therefore we simply present the normalized average power production for each row for all of the models considered. The CWBL and Jensen models applied to Horns Rev and Nysted field data use the parameters presented in Table \ref{table1}. We use a surface roughness length $\mathrm{z_{0,lo}}=0.002$ m to match the turbulence intensity at hub-height of $7.7\%$ used in the Horns Rev LES by Port\'e-Agel {\it et al.} \cite{por13}. The estimate is based on logarithmic laws for the mean (equation \eqref{Eq_profile_5}) and variance $\langle \overline{(u^{\prime+})^2} \rangle = B_\mathrm{1} - A_\mathrm{1} \log (z/\delta_\mathrm{H})$ in the inflow boundary layer with $A_\mathrm{1}\approx1.25$ and $B_\mathrm{1} \approx 1.6$ \cite{mar13,men13,ste14d}. This approach gives a value of $k_{\mathrm{w,0}}=0.0382$, which is used for all Jensen model results that are shown in figures \ref{figure4} to \ref{figure8}. For the thrust coefficient we use $\mathrm{C_T}=0.78$ for all cases, which is a characteristic average value of $\mathrm{C_T}$ for Horns Rev and Nysted turbines for wind speeds of around $8$ m/s \cite{por13,bar10b}. The field measurement data we compare to are reported for $5$ degree wind sectors \cite{bar09c,bar11} and therefore we average the model results over the same range of wind directions based on calculations at 0.5$^\circ$ intervals. In the first few rows of the entrance region of the wind-farm where turbines have not yet been subjected to wakes, the CWBL and Jensen model results are identical because the CWBL model reduces to the Jensen wake model there by construction. 

Figure \ref{figure4} shows that for the wind directions $265^\circ$ to $285^\circ$ the CWBL model compares reasonably well with the Horns Rev field measurements. The agreement of the Jensen model results compared to the field measurement data is similar to the agreement obtained with the CWBL model for $280^\circ$ and $285^\circ$. However, the Jensen model gives a larger overestimation of the wake effects for wind directions of $270^\circ$ and $275^\circ$. For the wind directions between $255^\circ$ and $260^\circ$ the agreement between the models and the Horns Rev data is less favorable. In particular, the CWBL and Jensen models both underpredict the wake effects. The figure shows that the stronger wake effects predicted by the CWBL model are in better agreement with the trends observed in the field data, but that the transition to the fully developed region appears to be too slow in the CWBL model. Figure \ref{figure4} also compares both model results with field measurement data for Nysted. Overall the agreement between the models and the Nysted data is better than for Horns Rev. However, the wake effects for the $268^\circ$, $273^\circ$, and $288^\circ$ wind directions are considerably underpredicted by both models, the predicted trends are, however, consistent with the field measurements and again the CWBL model does predict larger wake effects than the Jensen model for two of these three cases. It is also important to note that the $278^\circ$ direction is captured well by the CWBL model while the Jensen model greatly underestimates the power production for this wind direction.

Moreover, discrepancies between the field data and the model results are not only due to model limitations, but can also be attributed to limitations in the ability to specify the model inputs. Likewise, the experimental uncertainty includes uncertainty in conditions under which the field data have been obtained as well as subjective elements that arise during data analysis. In order to compare the analytical models with data under conditions and input variables that are better controlled than what is available from field data, we will compare with LES results for Horns Rev by Port\'e-Agel {\it et al.} \cite{por13} in the next section.

\section{Comparison with Horns Rev Large eddy simulations} \label{section_LES}

\begin{figure}[!h]
\centering
\includegraphics[width=0.99\textwidth]{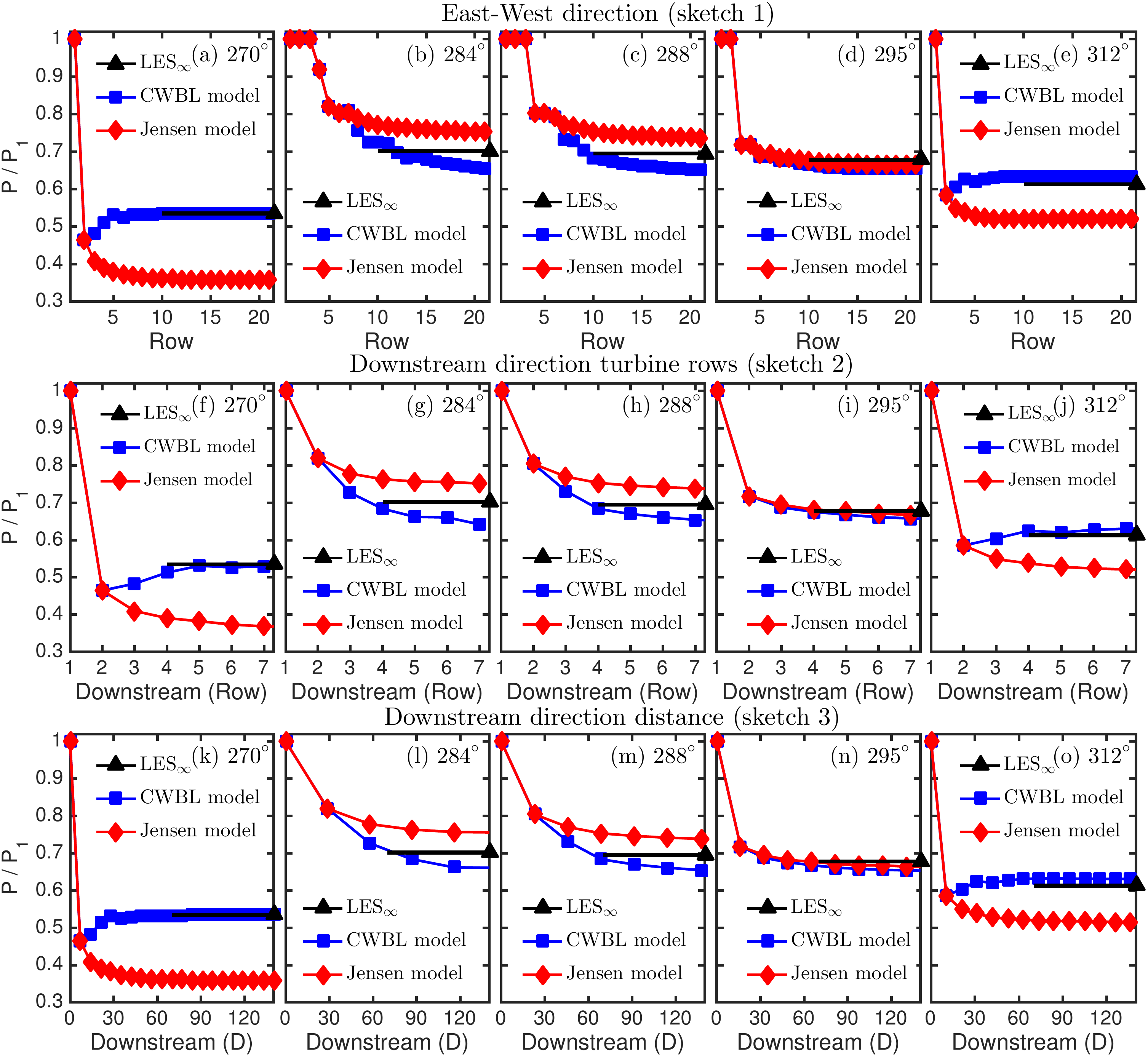}
\includegraphics[width=0.99\textwidth]{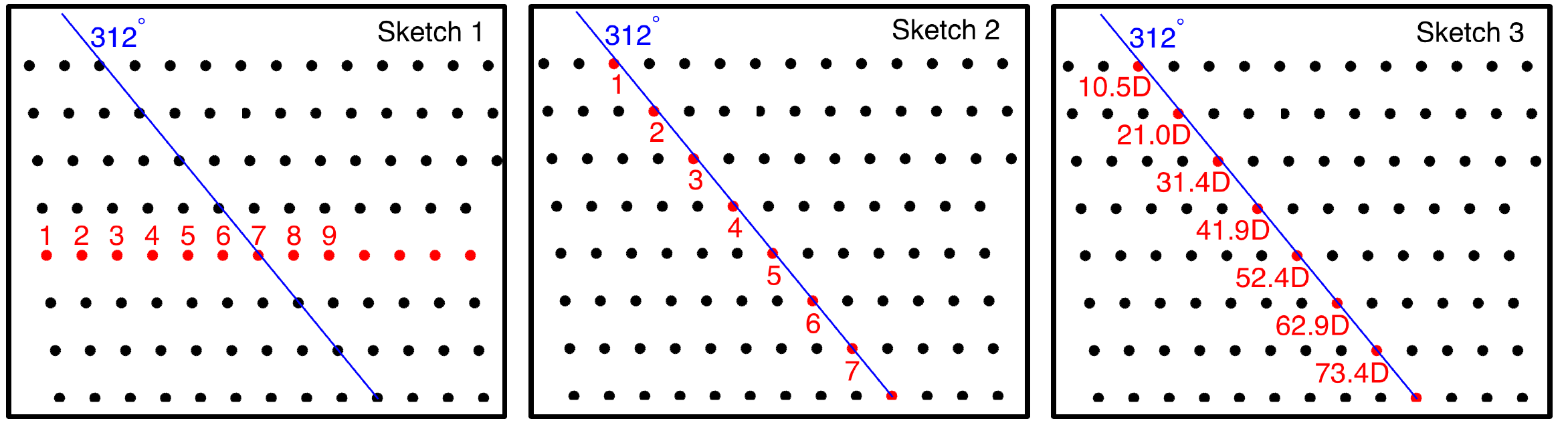}
\caption{CWBL and Jensen model results for an "extended Horns Rev" wind-farm with $25$ rows and $14$ columns compared to LES results from Port\'e-Agel {\it et al.} \cite{por13} in the fully developed region (triangles and horizontal line to show the asymptotic downstream behavior clearly, see details in text). (a-e) Results as function of row number, see sketch 1. (f-j) Results as function of row number in the downstream wind direction, see sketch 2. (k-o) Results as function of turbine distance in the downstream direction, see sketch 3. Note that the CWBL model predicts the LES value of the power output in the fully developed region much better than the Jensen model, especially for the $270^\circ$ and $312^\circ$ wind directions.}
\label{figure8}
\end{figure}

In this section the CWBL and Jensen models are compared with the Horns Rev simulations from Port\'e-Agel {\it et al.} \cite{por13}. The LES results span wind directions that range from $180-360^\circ$ and detailed results are presented for the wind directions $280^\circ$, $284^\circ$, $288^\circ$, $295^\circ$, and $312^\circ$. The comparisons show that the CWBL model captures the LES results very well. The wake losses at the second row, for which the CWBL and Jensen model give identical predictions, are overestimated for the $270^\circ$ and $312^\circ$ directions. For the $270^\circ$ and $312^\circ$ directions the power output for turbines further downstream is predicted better by the CWBL model than by the stand alone Jensen model due to the appropriate adjustment of the wake expansion coefficient $\mathrm{k_w}$ used in the Jensen model part of the CWBL model.

From visualizations of the normalized velocity at hub-height obtained from the CWBL model shown in figure \ref{figure6}, and the corresponding turbine power as function of the row number shown in figure \ref{figure6}, we see that for the $284^\circ$ and $288^\circ$ cases the turbines in the fourth row, see figure \ref{figure2}a, are the first ones that experience wake effects. As the nature of the wakes in the CWBL model depends on the number of wakes that interact with upstream turbines, see equation \eqref{equation_finitewindfarm}, the CWBL and Jensen model predictions only start to differ when wakes from the fourth turbine row reach turbines even further downstream, i.e.\ around the 7$^\mathrm{th}$ to 8$^\mathrm{th}$ row. From the visualizations in figure \ref{figure6} we see that for the $284^\circ$ and $288^\circ$ wind directions the wakes from the first row just miss the second row (especially for the $284^\circ$ direction), which means that slightly wider wakes (or smoother wake profiles) could have caused an earlier transition to the fully developed region than we see now. For the $284^\circ$ direction the turbines in the fourth row are, in addition, only partially in the wakes of the turbines from the first row. This means that a slightly wider wake would influence the predictions made by the CWBL model for that case. For such wind directions the use of more advanced wake models, see section 2.1, could help to further improve the predictions.

Figure \ref{figure7} shows the normalized total power output $\mathrm{P} / \mathrm{P_{max}}$ of Horns Rev as function of the incoming wind direction obtained from LES \cite{por13}, the CWBL model, and the Jensen model. The figure shows that both the CWBL and the Jensen model capture the main trends observed in the LES data. In agreement with the results presented in figure \ref{figure6} the CWBL model predictions are closer to the LES data for aligned wind directions, which are indicated by the vertical dashed lines, than the Jensen model results. However, for several intermediate wind directions the CWBL and Jensen model predictions are similar. The reason is that for these wind directions the power production of most turbines is not limited by wake effects. In addition, the first two rows in the prevailing wind direction of the wind-farm are modeled the same in the CWBL and Jensen model, so differences only become significant for wind-farms significantly larger than Horns Rev.

In order to illustrate the difference between the Jensen and CWBL model for larger wind-farms, figure \ref{figure8} compares both models for an ``extended Horns Rev" wind-farm with $25$ rows and $14$ columns. This figure depicts the power output for the different wind directions in three different ways. The middle and lower panels show the results in the downstream direction, see the sketch 1 in figure \ref{figure8}: the middle panels show the power output as function of the number of rows in the downstream direction, and the lower panels show the results as function of the downstream distance measured in turbine diameters, see sketches 2 and 3 in figure \ref{figure8}. Note that the horizontal range is the same for each wind direction to allow easier comparison of the results for the different wind directions. The figure reveals that the power output in the fully developed region predicted with the CWBL model is similar for the wind directions $284^\circ$, $288^\circ$, and $295^\circ$, and somewhat lower for the $270^\circ$ and $312^\circ$ wind directions when the distance between subsequent turbines in the downstream direction is small. This behavior seems to be in agreement with the LES data for the fully developed region (triangles and corresponding horizontal lines that indicate the asymptotic behavior). For these Horns Rev cases \cite{por13} we use the lowest power in the last three rows (measured in the wind direction) to estimate the power output in the fully developed region in order to minimize the effects due to limited time averaging. Figure \ref{figure8} also reveals that the Jensen model predicts ``marginal" wake effects for the $284^\circ$ and $288^\circ$ directions, while very strong wake effects are predicted for the aligned $(270^\circ)$ direction. These trends are not observed in the LES, see figure \ref{figure8}. For the $270^\circ$ and $312^\circ$ wind directions figure \ref{figure8}, shows that the CWBL model predictions agree very well with the LES observed behavior in the fully developed region compared to the stand alone Jensen model \cite{jen83,kat86} due to the adjusted wake expansion coefficient $\mathrm{k_w}$.

\section{Conclusions} \label{Section_discussion_conclusion}
In this paper we have introduced a generalized CWBL model that enables analysis of arbitrary wind directions, compared to the original model presented in Ref. \cite{ste14g}, and applications to fairly general wind turbine geometric arrangements within a wind-farm. We then compared the (generalized) CWBL and Jensen models with field measurements. We also provided comparisons with LES results in order to validate the CWBL model in a setting where the key control parameters are fully defined. In general, the agreement between LES and the models is better than the agreement between the models and the field measurement data. In the entrance region of the wind-farm the CWBL model reduces to the wake (Jensen) model and therefore both models give the same results in that region. Further downstream the CWBL model is different than the wake (Jensen) model due to the coupling with the ``top-down" (Calaf {\it et al.}) model, which captures the interaction between the wind-farm and the ABL. Comparisons with LES and field measurements reveal that the CWBL model gives improved predictions for the power output in the fully developed region of the wind-farm compared to the Jensen model. We find that the CWBL model compares particularly well with the LES results from Port\'e-Agel \cite{por13}, which have been obtained under neutral atmospheric conditions and for a constant wind direction. Such better agreement is perhaps not unexpected since these are the same conditions assumed in the CWBL model. However, we note that for some particular wind directions the CWBL model underpredicts the rate at which the transition to the fully developed region occurs. Overall we conclude that the CWBL model shows considerable promise but further studies will be necessary to extend this modeling approach to more general conditions and to define the uncertainty in the predictions.

\section{Acknowledgments:}
The authors are grateful for fruitful interactions and discussions with F. Port\'e-Agel at the beginning of this work. RJAMS's work is supported by the program `Fellowships for Young Energy Scientists' (YES!) of the Foundation for Fundamental Research on Matter (FOM), which is financially supported by the Netherlands Organization for Scientific Research (NWO). Partial support has also been provided by the US National Science Foundation from grant IIA 1243482 (the WINDINSPIRE project). We thank the anonymous referees for their valuable comments during the referee process.

\FloatBarrier

\section*{Appendix} \label{Appendix}

\begin{figure}
\subfigure{\includegraphics[width=0.99\textwidth]{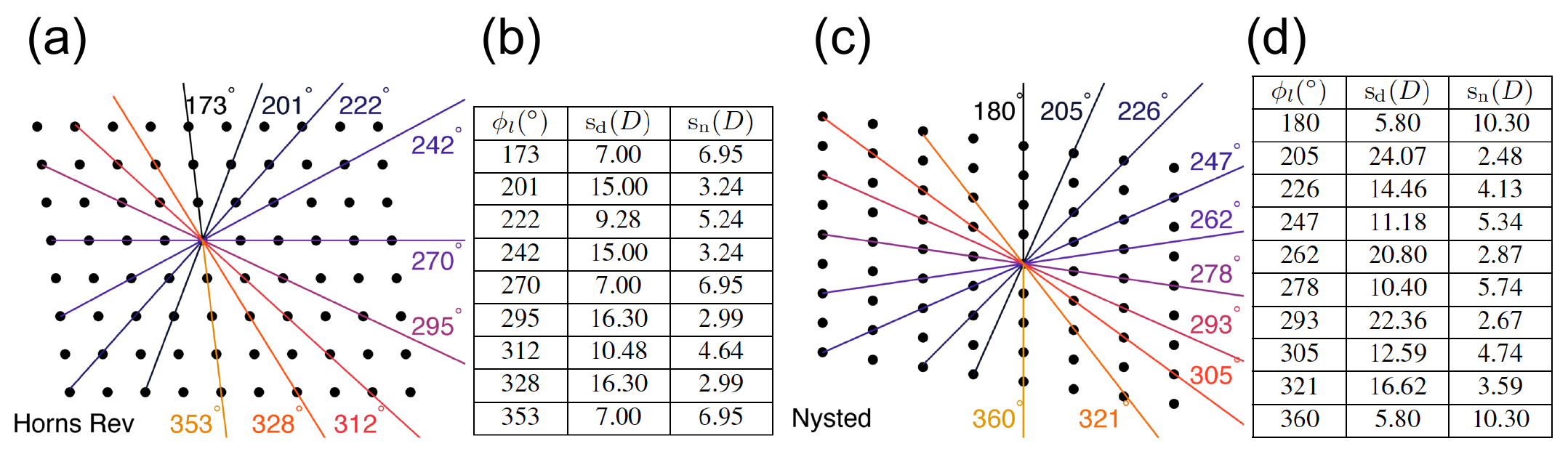}}
\centering
\caption{Panels (a) and (c) show the main symmetries for Horns Rev and Nysted, respectively. The tables (b) and (d) indicate the distance between the turbines in the downwind direction, $\mathrm{s_d}$, and the corresponding distance between neighboring rows, $\mathrm{s_n}=\mathrm{s_x}\mathrm{s_y}/\mathrm{s_d}$.}
\label{figure10}
\end{figure}

\begin{figure}
\centering
\subfigure{\includegraphics[width=0.99\textwidth]{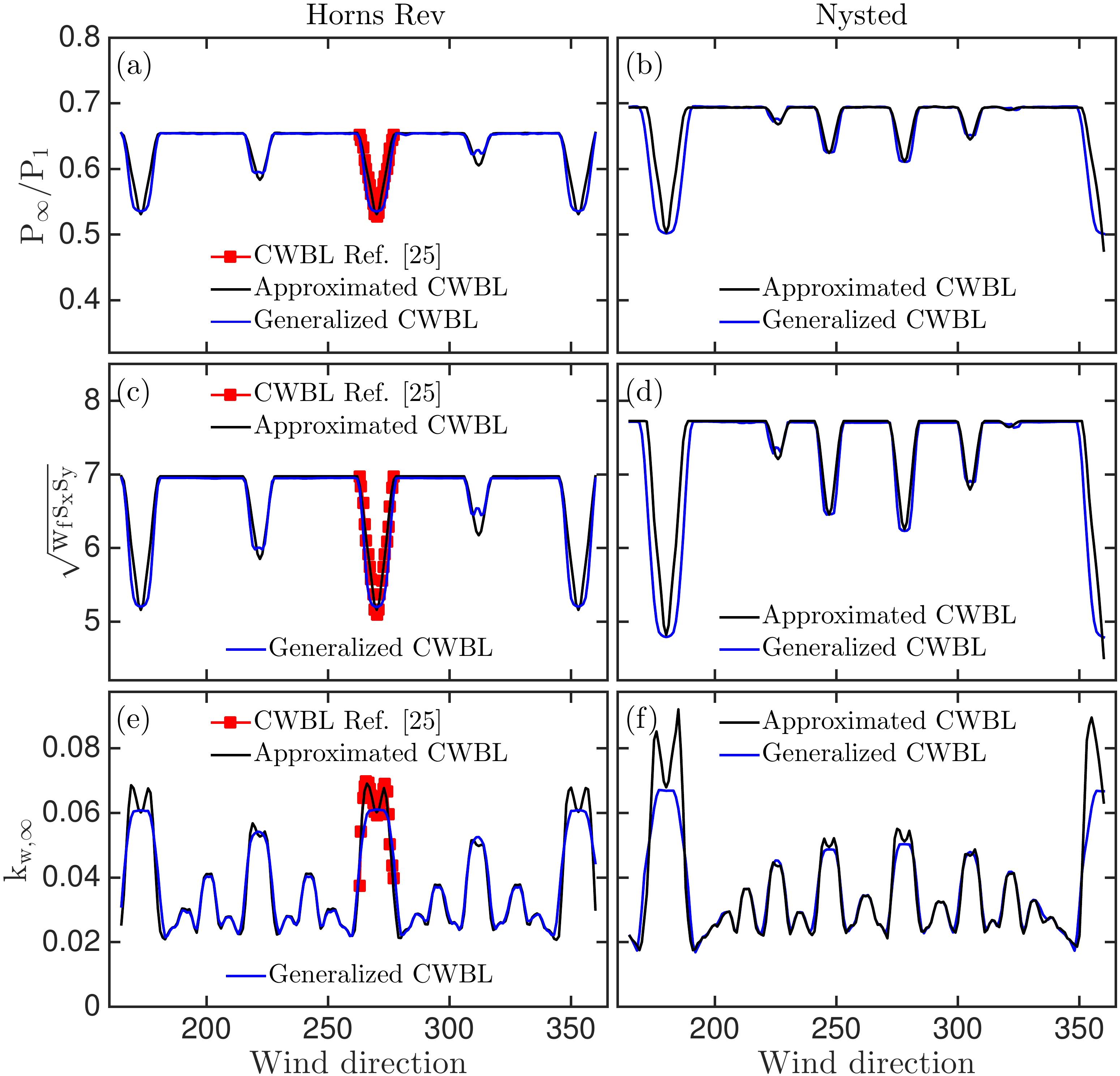}}
\caption{Detailed tests of the calibration of velocity threshold chosen for the generalized CWBL model. The figure shows that results obtained with the new generalized CWBL calculations (blue lines) for Horns Rev and Nysted agree quite well with the method underlying the original CWBL model \cite{ste14g} (but applied here to other inflow directions). Also, good agreement exists with the approximated CWBL model presented in this appendix. (a,b): Normalized power output in the fully developed region. (c,d): Effective geometric mean turbine spacing $\sqrt{\mathrm{w_f}\mathrm{s_x}\mathrm{s_y}}$. (e,f): Wake expansion coefficient $\mathrm{k_{w,\infty}}$ in the fully developed region. Data are averaged over $3^\circ$ bin sectors (instead of the coarser $5^\circ$ bin sectors used in the main sections) to compare the model results in detail.}
\label{figure11}
\end{figure}

In the CWBL approach introduced by Stevens {\it et al.} \cite{ste14g} the coupling between the Jensen wake model \cite{jen83,kat86} and the Calaf {\it et al.} \cite{cal10} ``top-down" model was obtained by determining what was called the effective spanwise spacing $\mathrm{s_{ye}}$. While accurate and detailed, the version of the CWBL model in Ref.\ \cite{ste14g} was limited to particular inflow angles for which one could easily define the streamwise and spanwise spacings. However for general angles and wind-farm geometries, it becomes very difficult to define $\mathrm{s_x}$ and $\mathrm{s_y}$ or $\mathrm{s_{ye}}$. In the present appendix we calibrate the velocity threshold used in the CWBL model introduced in section 2.3 so that predictions obtained here and in Ref.\ \cite{ste14g} agree in the cases where both versions can be applied. Once calibrated, the method used in section 2.3 can be applied to general wind-farm layouts. In the generalized CWBL method introduced in section \ref{section_coupling} the coupling is obtained by determining the effective wake area coverage fraction $\mathrm{w_f}$. In the ``top-down" model part of the model the coupling occurs through the term $ (\pi \mathrm{C_T}) / (8 \mathrm{w_f}\mathrm{s_x}\mathrm{s_y})$ in equation \ref{Eq_defz0hi}. This term is evaluated by determining the wake coverage area in the fully developed region and is related to the effective spanwise spacing used in Ref.\ \cite{ste14g} by
\begin{equation}
 \mathrm{w_f}\mathrm{s_x}\mathrm{s_y} D^2 = \mathrm{s_x}\mathrm{s_y} D^2 - \mathrm{A_{ch}} = \mathrm{s_x} \mathrm{s_{ye}} D^2,
\end{equation}
where $\mathrm{A_{ch}}$ is the area of the ``high speed channels". These high speed channels should be excluded from the effective control volume that is used in the Calaf {\it et al.} ``top-down" model as these ``high speed channels" do not participate in the vertical momentum and kinetic energy vertical mixing as envisioned in the ``top-down" model. In order to use the generalized CWBL model one needs a threshold value to determine when a particular area should be identified as wake. This is done in this appendix. Figure \ref{figure11} shows that when a $5\%$ threshold is used, such that a wake is defined as the region where $\mathrm{\mathrm{\overline{u}}(x,y)} < 0.95 \mathrm{\overline{u}_0}$, the resulting predictions agree well with the results obtained with the original CWBL model in which the effective spanwise spacing is used. The remainder of the appendix gives additional details of what is shown in figure \ref{figure11}.

In the original CWBL method \cite{ste14g} the effective spanwise spacing $\mathrm{s_{ye}}$ for an aligned wind-farm was determined by calculating the smallest spanwise distance for which the effect of turbine wakes originating from laterally adjacent turbine columns on the power production in the fully developed region is less than $1\%$ \cite{ste14g}. That work showed (using the Jensen model) that the effective spanwise spacing $\mathrm{s_{ye}}$ is relatively independent of $\mathrm{k_{w,\infty}}$ and $\mathrm{s_x}$, and can be approximated by the constant value $\mathrm{s_y}^*=\mathrm{s_{ye}}=3.5$ for aligned wind-farms. The procedure to determine the effective spanwise spacing $\mathrm{s_{ye}}$ for more general wind turbine configurations or wind directions is difficult and computationally expensive. It is instructive to provide a comparison between the generalized CWBL approach introduced in section \ref{section_coupling} and the effective spanwise spacing based approach for some particular inflow angles in which the original method can still be applied. Specifically, a good approximation for the effective spanwise spacing method can be obtained by first determining the main alignments for which the effective spanwise spacing $\mathrm{s_{ye}}$ is known. This approximation is introduced here to allow comparison of the generalized approach presented in section \ref{section_coupling} over a much wider range of wind directions. For Horns Rev and Nysted this is done in figure \ref{figure10}. Figures \ref{figure10}b and \ref{figure10}d indicate the distance between consecutive turbines in the downstream direction for each of these wind directions, $\mathrm{s_d}$, and the corresponding distance between neighboring turbine rows, $\mathrm{s_n}=\mathrm{s_x}\mathrm{s_y}/\mathrm{s_d}$. Here $\mathrm{s_x}$ and $\mathrm{s_y}$ are the nondimensional streamwise and spanwise distance between the turbines for the aligned configuration, i.e.\ the $270^\circ$ direction for Horns Rev and the $278^\circ$ direction for Nysted. 

An approximated (original) CWBL model, involving calculation of the effective spanwise spacing $\mathrm{s_{ye}}$, is then obtained by using the following iterative procedure:\\

\noindent Step 1: Same as step 1 shown in section \ref{section_coupling}. 
In addition, for a given wind-farm configuration, a set of angles $\phi_l$ and corresponding streamwise and spanwise spacings must be determined and tabulated. Specifically, $\phi_l$ ($l=1,2,3...$) are a set of particular lattice angles that align with the array structure of the wind-farm.
For example, in a wind-farm with a square array, i.e. $\mathrm{s_x}=\mathrm{s_y}$ when $\phi_\mathrm{wind}=0$, then $\phi_l=45^o$ is an angle for which the wind-farm appears again as a regular array, but now with streamwise and spanwise spacings increased and decreased by factors $\sqrt{2}$ and $1/\sqrt{2}$, respectively. Figure \ref{figure10} provides an example of such a tabulated set for the Horns Rev and Nysted wind-farms of values of special angles and spacings for which $s_n\lesssim 3$.
\\
\\
Step 2: Determine $\psi$, i.e.\ the angle difference between the actual wind direction and the closest wind direction that aligns with the geometry of the wind-farm (previously tabulated, see step 1 and figure \ref{figure10}), using
\begin{equation}
\psi=| \phi_\mathrm{wind}-\phi_l |,
\end{equation}
where $\phi_\mathrm{wind}$ is the wind direction and $\phi_l$ is the closest tabulated wind direction angle that aligns with the wind-farm lattice. 
\\
\\
Step 3: For the current value of $\mathrm{k_{w,\infty}}$ determine the angle $\theta$
\begin{equation}
\label{equationstep2}
	\theta=\mathrm{arctan}\left(\frac{0.5+\mathrm{k_{w,\infty}}\mathrm{s_d} (\phi_\mathrm{l})}{\mathrm{s_d} (\phi_l)}\right),
\end{equation}
which indicates the angle (in radians) needed for an expanding wake to reach the center of each of the turbines in the next row and $\mathrm{s_d} (\phi_l)$ is the downstream distance between consecutive turbine rows for the closest wind direction that aligns with the wind-farm geometry. The parameter $0.5$ was obtained from a comparison with the full CWBL model calculations, see figure \ref{figure11}.\\
\\
Step 4: Determine $\mathrm{s_{ne}}$ as a linear interpolation between the $\mathrm{s_y}^\mathrm{*} = 3.5$ empirical value known to be appropriate for aligned wind-farms, and the geometric spanwise spacing of the ``closest aligned" configuration $\mathrm{s_n}(\phi_l)$ determined above:
\begin{equation} 
\mathrm{s_{ne}} = \min\left[ \mathrm{s_n} (\phi_l)+(\mathrm{s_n} (\phi_l) -\mathrm{s_y}^*)( 1 - \mathrm{min}[\psi/\theta,1]), \mathrm{s_n} (\phi_l) \right]
\end{equation}
In addition, we demand that $\mathrm{s_{ne}} \leq \mathrm{s_n} (\phi_l)$. This interpolation approach enables us to obtain smooth results as function of inflow wind direction, but is quite complicated to implement in practice.\\
\\
Step 5: Calculate $\langle \mathrm{\overline{u}} \rangle(\mathrm{z_h}) / \langle \mathrm{\overline{u}_0} \rangle(\mathrm{z_h})$ using equation \eqref{Eq_veloc} with the ``top-down'' model using $\mathrm{s_d}$ and $\mathrm{s_{ne}}$ and find the wake expansion $\mathrm{k_{w,\infty}}$ that makes it consistent with the wake (Jensen) model. Note that $\mathrm{s_d}$ and $\mathrm{s_{ne}}$ correspond to $\mathrm{s_x}$ and $\mathrm{s_y}$ for the $270^\circ$ direction for Horns Rev and to the $278^\circ$ direction for Nysted.\\
\\
Steps 2 to 5 are iterated until the turbine velocity in the wake (Jensen) model and the ``top-down" part of the CWBL model are converged and agree among each other.\\

Figure \ref{figure11} shows a comparison between the original \cite{ste14g}, the approximated (original), and generalized CWBL model. Here we note that the approximated CWBL model neglects the dependence of $\mathrm{s_y}^*$ on the wake expansion coefficient $\mathrm{k_{w,\infty}}$ and $\mathrm{s_x}$ \cite{ste14g}. The figure shows that the effective geometric mean turbine spacing $\sqrt{\mathrm{w_f}\mathrm{s_x}\mathrm{s_y}}$ obtained with the approximated procedure agrees well with the full model calculations \cite{ste14g}. We note that the full model calculations (neglecting the effect of small angles with respect to the aligned configuration) have been used to set the constant ``0.5" in step 3 (equation \eqref{equationstep2}). The benefit of the approximated (original) CWBL procedure is that it can be used more easily to check the consistency of the results from the new generalized CWBL model presented in this work for all wind directions than the full original calculations. This is done in figure \ref{figure11}, which shows that the power production in the fully developed region (top panels), the effective geometric mean turbine spacing (middle panels), and the wake expansion coefficient in the fully developed region (lower panels).

We can conclude that using the new generalized method, outlined in section \ref{section_coupling}, approximates the original model results closely, when the $5\%$ velocity threshold mentioned above is used. The figures show that the ``wake width" around the aligned wind directions is slightly larger using the new method than with the original method. Figure \ref{figure11} shows that this effect is most pronounced, but still relatively small, when the distance between consecutive downstream turbines is small, i.e.\ see the Nysted wind directions around 180$^\circ$, and leads to corresponding difference in the predicted wake expansion coefficient in the fully developed region $\mathrm{k_{w,\infty}}$. It is important to note that the differences in the predicted $\mathrm{k_{w,\infty}}$ only have a relatively small influence on the results as is indicated by the difference in the predicted power production in the fully developed region. Therefore, in practical applications the generalized method in section \ref{section_coupling} is the method of choice.

\end{document}